\documentclass[%
 aip,
 amsmath,amssymb,
reprint,%
]{revtex4-1}

\usepackage{dcolumn}
\usepackage{bm}
\usepackage{amsmath}
\usepackage{braket}
\usepackage{graphicx}
\usepackage{float}

\usepackage[utf8]{inputenc}
\usepackage[T1]{fontenc}
\usepackage{mathptmx}
\usepackage{etoolbox}
\usepackage[normalem]{ulem}
\newcommand{\daria}[1]{\textcolor{black}{{#1}}}

\usepackage{xcolor}

\makeatletter
\def\@email#1#2{%
 \endgroup
 \patchcmd{\titleblock@produce}
  {\frontmatter@RRAPformat}
  {\frontmatter@RRAPformat{\produce@RRAP{Authors to whom correspondence should be addressed: \href{mailto:#2}{#2, david.wisbey@slu.edu, dariakowsari@wustl.edu}}}\frontmatter@RRAPformat}
  {}{}
}%
\makeatother

\begin{document}

\preprint{APL_eBeam_Nb_v1.08}

\title{Fabrication and surface treatment of electron-beam evaporated niobium for low-loss coplanar waveguide resonators}

\author{D. Kowsari}
\author{K. Zheng}
\author{J. T. Monroe}
\affiliation{Department of Physics, Washington University, Saint Louis, Missouri 63130, USA.}

\author{N. J. Thobaben}
\affiliation{Department of Physics, Saint Louis University, Saint Louis, Missouri 63103, USA.}

\author{X. Du}

\affiliation{Department of Physics, Washington University, Saint Louis, Missouri 63130, USA.}

\author{P. M. Harrington}
\affiliation{Department of Physics, Washington University, Saint Louis, Missouri 63130, USA.}

\affiliation{Research Laboratory  of Electronics, Massachusetts Institute of Technology, Cambridge, Massachusetts 02139, USA.}

\author{E. A. Henriksen}
\affiliation{Department of Physics, Washington University, Saint Louis, Missouri 63130, USA.}
\affiliation{Institute for Materials Science and Engineering, Washington University, Saint Louis, Missouri 63130, USA.}

\author{D. S. Wisbey}
\affiliation{Department of Physics, Saint Louis University, Saint Louis, Missouri 63103, USA.}

\author{K. W. Murch}
\email{murch@physics.wustl.edu}
\affiliation{Department of Physics, Washington University, Saint Louis, Missouri 63130, USA.}

\date{\today}

\begin{abstract}
We characterize low-loss electron-beam evaporated niobium thin films deposited under ultra-high vacuum conditions. Slow deposition yields films with a high superconducting transition temperature ($9.20 \pm 0.06 \rm ~K$) as well as a residual resistivity ratio of $4.8$. We fabricate the films into coplanar waveguide  resonators to extract the intrinsic loss due to the presence of two-level-system fluctuators using microwave measurements. For a coplanar waveguide resonator gap of $2~\mu \rm m$, the films exhibit filling-factor-adjusted two-level-system loss tangents as low as $1.5 \times 10^{-7}$ with single-photon regime internal quality factors in excess of one million after removing native surface oxides of the niobium.     
\end{abstract}

\maketitle
Recent achievements in quantum computing have {shown} that superconducting circuits are one of the most promising platforms to realize the long-sought challenge of building a fault-tolerant quantum computer \cite{Arute2019,Jurcevic2021,Wu2021}. The performance of such devices are, however, limited by decoherence sources such as quasiparticles \cite{DeVisser2014,Barends2011,Corcoles2011}, magnetic vortices \cite{Song2009}, and radiation effects\cite{Sage2011}. Recent advances in fabrication techniques and microwave engineering have significantly reduced the impacts of the above-mentioned defects \cite{Kreikebaum2016,Sandberg2013,Chiaro2016}, thereby leaving two-level-system (TLS) fluctuators as the most prominent source of loss in superconducting circuits \cite{Lisenfeld2019,DeGraaf2020,Niepce2020}. It has been shown that TLS defects are mainly located at metal-air (MA), substrate-air (SA), and metal-substrate (MS) interfaces \cite{Gao2008,Wisbey2010,Bilmes2020,Grunhaupt2017,Calusine2018}. The contribution of losses from these interfaces can be minimized by implementing careful surface treatments thus enhancing the coherence of the devices \cite{Place2021,Monroe2021}.

Owing to its high superconducting transition temperature, critical field, and low microwave loss \cite{Halbritter1987}, niobium (Nb) has become one of the common materials used in the fabrication of superconducting circuits \cite{Nersisyan2019,Blok2021}. Nevertheless, the known stoichiometric range of its native oxides results in a complex loss-inducing MA interface \cite{Bach2006,Bach2009_Thesis}. Lately, it has been shown that removing the oxides on the MA interface of Nb films can result in highly coherent devices \cite{Altoe2020, Verjauw2021}. \daria{Additionally}, conventional niobium deposition techniques, such as DC magnetron sputtering can result in a damaged MS interface due to the presence of high energy argon ions \cite{Siemers2014} and point defects stemming from trapped argon atoms \cite{DHeurle1970}.

In order to prevent these sources of decoherence, here we investigate a refined deposition method \cite{Morohashi2001} of Nb utilizing an ultra-high vacuum (UHV) electron-beam evaporator. We fabricate the Nb thin films into coplanar waveguide (CPW) resonators, which are well-known for their ease of fabrication as well as their sensitivity to the true intrinsic TLS defect density of the materials \cite{Gao2008,Pappas2011}. \daria{We observe that surface treatment results in lower TLS densities compared to prior studies using the same CPW geometry optimized for sensitivity to TLS defects\cite{McRae2020,Gao2008,Sage2011,Chang_2013}.} This study further establishes Nb, with appropriate surface treatment, as an ideal material for the fabrication of highly coherent superconducting qubit processors.

Samples are fabricated on a 2-inch, (100)-oriented, high resistivity ($> 8~\rm{k}\Omega\cdot\rm{cm}$), single-side polished intrinsic silicon substrate cleaned in a Piranha solution (3:1 mixture of sulfuric acid and hydrogen peroxide) at $120~^{\circ}\rm C$ for $10$ minutes followed by a $5$-minute etch in a buffered-oxide-etch (BOE) solution to remove organic contaminants as well as the native silicon surface oxide \cite{Place2021,Zeng2015}. The BOE solution is a 6:1 mixture of ammonium fluoride ($\rm{NH_{4}F}$) and hydrofluoric acid ($\rm{HF}$). The substrate is then pumped down in a UHV electron-beam evaporator (AJA ATC-ORION-8E) with a base pressure lower than $5~\rm{nTorr}$.

After loading the wafer, a $200$~nm layer of 99.95\% purity Nb is evaporated onto the substrate at a rate of $1.2$~nm/min, which is commensurate to the previous study\cite{Morohashi2001}. Note that the substrate is inevitably heated during this process due to the high melting point of Nb. Since pure Nb quickly adsorbs impurities \cite{Halbritter1987}, especially when heated, we let the sample remain under UHV conditions to cool down for 1-2~hours prior to proceeding to the next steps. 

We spin and softbake the Nb samples with a high resolution photoresist (MicroChem S1805) and pattern the coated wafer with a Heidelberg DWL $66+$ photolithography system. The pattern consists of $8$ hanger-style, quarter-wavelength CPW resonators with a gap (width) of $2~\mu \rm m$ ($3~\mu \rm m$).  Devices are simulated to have frequencies ranging from $5.2$ to $7$ GHz with coupling quality factors ${\sim}6 \times 10^5$ (simulated using SONNET microwave software). We develop the exposed resist by using a metal-ion-free solution (MicroChem MF-319). A reactive ion etch system (Oxford Plasmalab 100) is then used to etch the samples using a fluorine chemistry ($\rm{SF_6}$). To assist the removal of the residual resist, the samples are first ashed for 30 seconds using oxygen plasma (Plasma Etch PE 50, 100 W, 15 cc/min) and then soaked in N-Methyl-2-pyrrolidone (NMP) heated to $70~ ^{\circ} \rm C$ for 8 hours. At last, samples are coated with the photoresist (S1805) to enhance their preservation over time\cite{Altoe2020} and protect against damage caused by dicing.

\begin{figure}[t]
    \centering
    \includegraphics[width=8.5 cm]{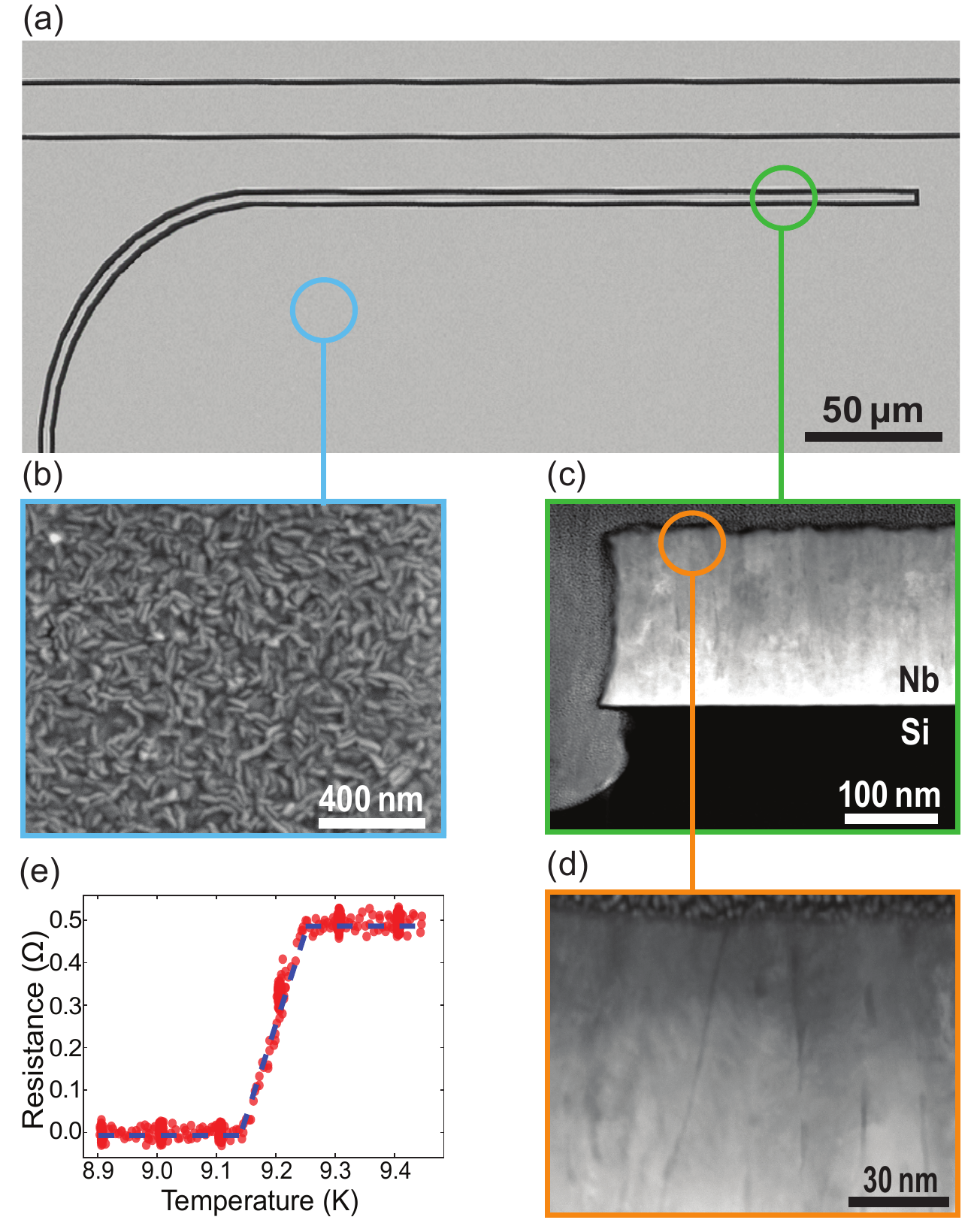}
    \caption{Surface {topography} of the fabricated CPWs. (a) SEM image shows the fabricated resonator coupled to the feedline with a CPW gap of $2 ~\mu \rm m$. (b) Detailed SEM image displays the compact elongated niobium grains formed on the surface. (c) A dark-field STEM cross-sectional image illustrates the anisotropic etch. (d) A STEM cross-sectional image depicts grain sizes exceeding $20 ~\rm nm$.  (e) The four-probe resistance measurement shows a superconducting transition temperature $\rm{T_c}=9.20 \pm 0.06 \rm ~K$. \daria{The uncertainty in the transition temperature is taken as the full temperature range of the transition.   The RRR is calculated by dividing the surface resistance at $310~\rm{K}$ to that right before the transition at $9.258~\rm{K}$ as $2.33~\Omega/0.49~\Omega\simeq4.8$ exhibiting high quality of the films.}}
    \label{Fig1}
\end{figure}

\begin{figure}[b]
    \centering
    \includegraphics[width=8.5 cm]{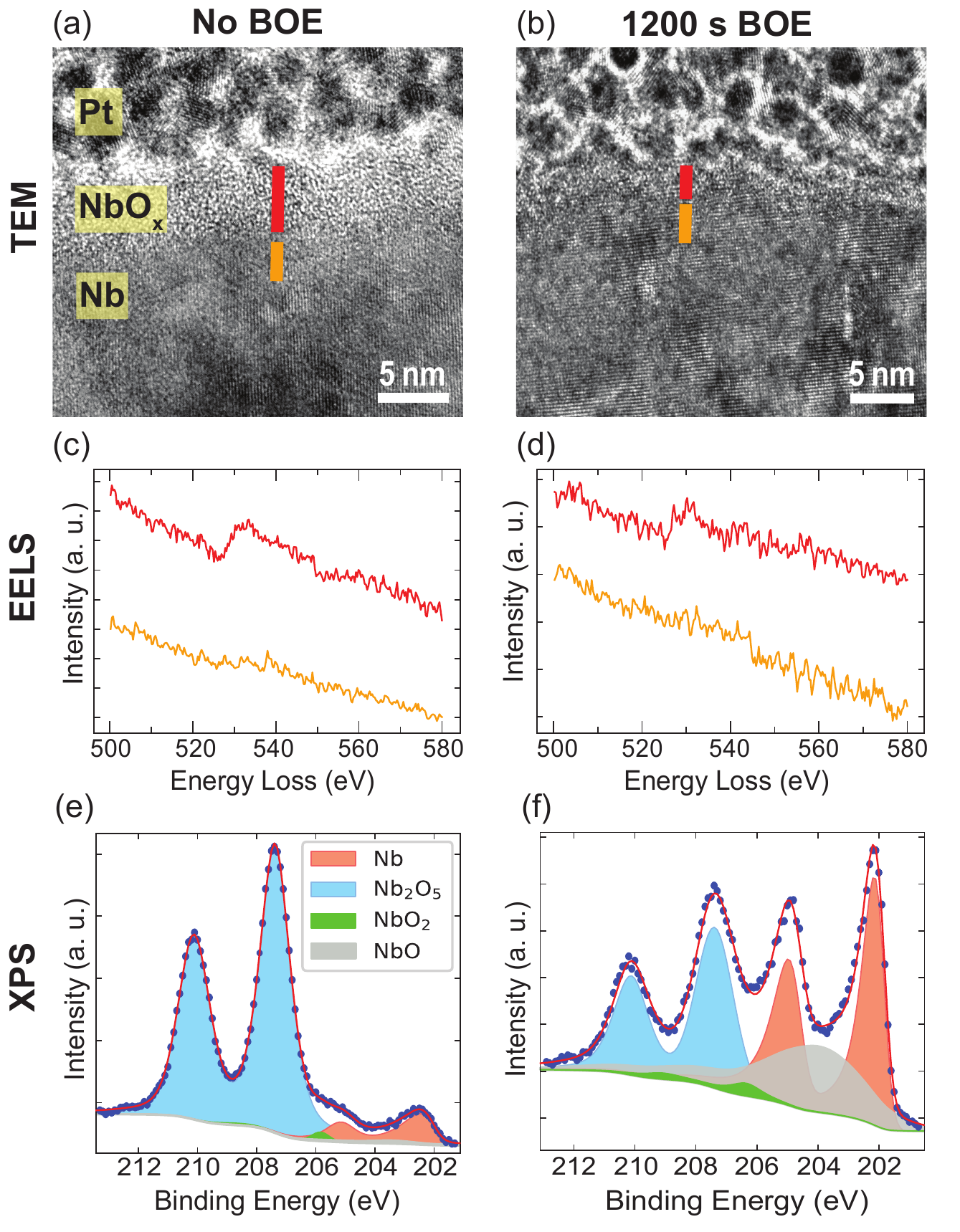}
    \caption{Comparison of the metal-air interface characteristics with different cleaning schemes. \daria{TEM samples are prepared using a focused ion beam after the deposition of a platinum (Pt) protective layer on top.} TEM images show (a) a clear band of $\rm NbO_x$ ${\sim}5 \rm~nm$ thick on top of the Nb surface and (b) after cleaning the sample for 20 minutes with BOE, the oxide layer thickness was reduced to ${\sim}1.7 \rm~nm$. Red and orange lines in (a, b) indicate regions where averaged EELS spectra are collected. (c, d) The averaged EELS spectra with orange and red traces corresponding to regions depicted in (a, b). The spectra in the oxide region (red) displays a significant peak around 535 eV corresponding to oxygen. This peak is absent along the Nb metal (orange), showing that there is no trace of oxygen in the Nb film. (e) $\rm{Nb_{3d}}$ XPS measurements of the Nb surface show different niobium-oxygen compounds, dominated mostly by $\rm{Nb_{2}O_5}$ oxide prior to treating the surface.  (f) After surface cleaning, we observe a clear increase in the Nb peak relative to the oxides.} 
    \label{Fig2}
\end{figure}

Figure~\ref{Fig1} illustrates the surface {topography} of the Nb films fabricated into CPWs (Fig.~\ref{Fig1}(a)) using the \daria{above-mentioned} techniques. The scanning electron microscopy (SEM) image shows the elongated Nb grains formed on the surface (Fig.~\ref{Fig1}(b)). The dark-field scanning transmission electron microscopy (STEM) cross-sectional image of the etched films shows that our etching process is anisotropic (Fig.~\ref{Fig1}(c)). In this study, the etching has been engineered to result in trenches shallower than $500~ \rm nm$ for all the fabricated devices to maintain the effective substrate dielectric constant in the devices, and thereby  avoid deviations from the desired resonance frequencies \cite{McRae2020}. The STEM image shows grain sizes greater than $20~\rm{nm}$ (Fig.~\ref{Fig1}(d)) exceeding previous values reported using the same deposition technique \cite{Morohashi2001}. Correlation between the grain sizes and materials loss has been recently studied \cite{Premkumar2021}, substantiating the advantage of larger grain sizes by treating the grain boundaries as Josephson weak-links \cite{Bonin1991}. Finally, the films resulted in a superconducting transition temperature $\rm {T_c} = 9.20 \pm 0.06 \rm ~K$ and a residual resistivity ratio (RRR) of 4.8 (Fig.~\ref{Fig1}(e)) exhibiting the high quality of the evaporated Nb, in accord with recent results with sputtered Nb\cite{Premkumar2021}. \daria{Note that the films deposited at higher pressures ($>~6~\rm{nTorr}$) resulted in a significantly lower transition temperature $\rm{T_c}\simeq7.85~\rm{K}$ as well as a RRR of only $1.8$, which can be attributed to the fact that having a low deposition rate makes the quality of the films extremely sensitive to the deposition environment. Therefore, having a UHV deposition environment is a crucial condition for achieving high quality films using this technique.}

Transmission electron microscopy (TEM) images show a clear band of about $5~\rm{nm}$ of oxide on top of the Nb, which is reduced to ${\sim}1.7~\rm{nm}$ after removing the surface oxides in a BOE solution for 20 minutes (Fig.~\ref{Fig2}(a, b)). Note that it takes about 20 minutes to transfer the sample to the characterization instruments, during which the oxide grows back following Cabrera-Mott theory \cite{Verjauw2021}. To identify the elements present in the films, we perform electron-energy loss spectroscopy (EELS) measurements. Averaged EELS spectra are displayed in Fig.~\ref{Fig2}(c, d) along the red (orange) lines  for the oxide (metal) regions of the TEM images in Fig.~\ref{Fig2}(a, b). The regions indicated by red lines exhibit a clear peak located at ${\sim}535~\rm{eV}$, which corresponds to the presence of oxygen. \daria{Below this $\rm{NbO_x}$ band, there is no sign of this peak. Furthermore, EELS data revealed a clear reduction in the average oxygen content from ${\sim}50\%$ in the oxide band to less than $5\%$ while scanning inside the metallic Nb film indicating the absence of oxygen impurities in the films.}

\begin{figure}[t]
    \centering
    \includegraphics[width=8.5 cm]{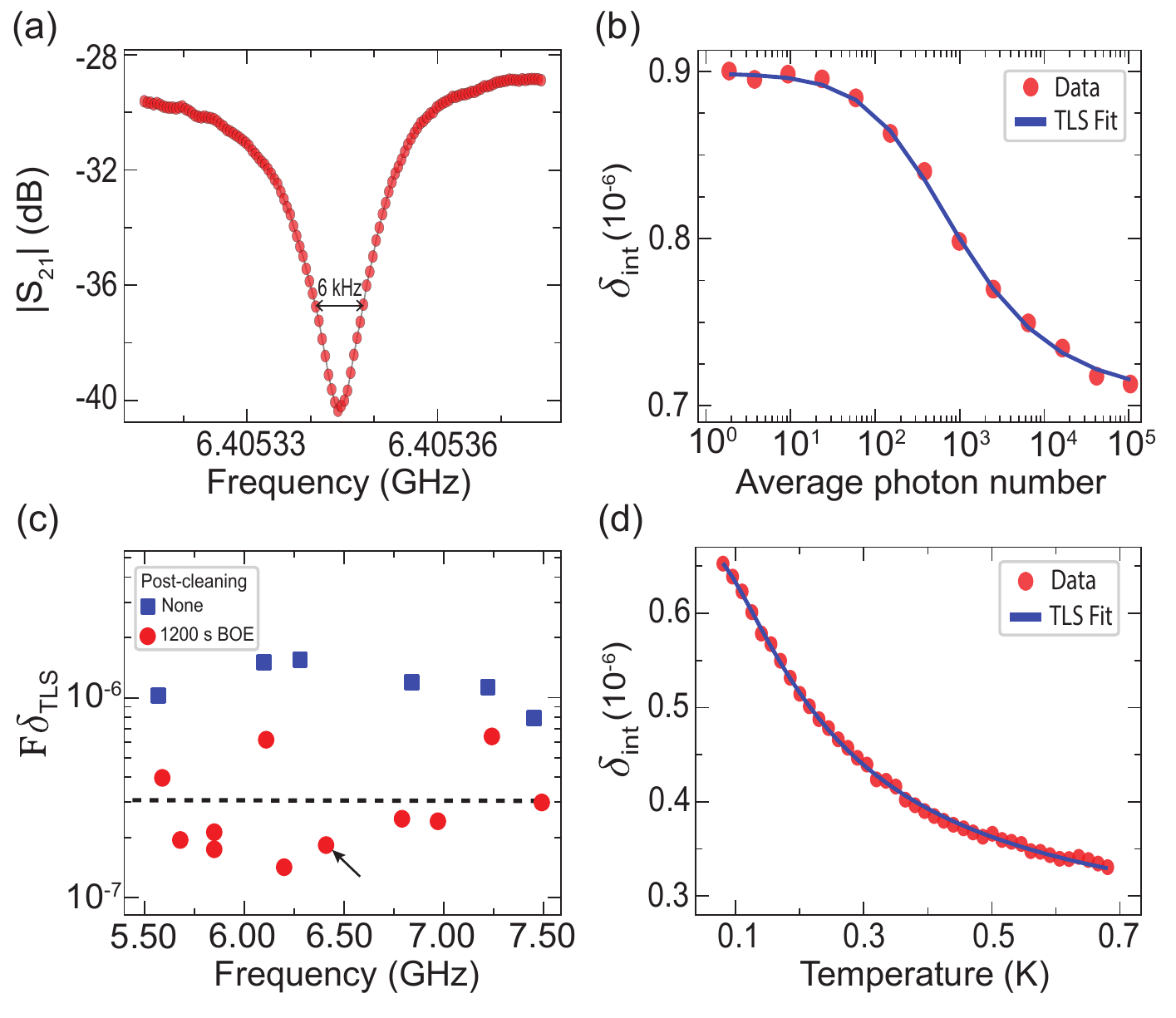}
    \caption{Microwave measurements of the CPW resonators. (a) Transmission profile near the resonance at single-photon power shows a full width at half maximum of 6 kHz corresponding to an internal quality factor of $1.06\times 10^6$ for the device indicated by the arrow in (c). (b) The internal loss of the same device as a function of the average photon number, showing minimal dependence on power. Data points were fit using the TLS model (blue solid line), resulting in \daria{$\rm{F}\delta_{TLS}=(1.94\pm0.02)\times 10^{-7}$}. (c) Comparison of the extracted F$\delta_{\rm{TLS}}$ of the fabricated CPW resonators from the TLS fit of the power scans as a function of the resonance frequencies across three chips for different post-cleaning treatments. Red circles indicate results for devices that were etched with a BOE solution for 1200 seconds before placement into a cryogenic vacuum environment for measurement.  Blue squares correspond to results for samples that did not receive this pre-measurement etch treatment. The dashed line indicates the average F$\delta_{\rm{TLS}}$ value of the treated devices. (d) TLS fit of the temperature scan \daria{at high circulating power ($-93~\rm{dBm}$), corresponding to an average photon number of ${\sim}10^{5}$, resulting in $\rm{F}\delta_{TLS}=(4.39\pm0.02)\times 10^{-7}$}, showing a good agreement with the power scan result in (b).}
    \label{Fig3}
\end{figure}

We employ X-ray photoelectron spectroscopy (XPS) to further investigate the  compounds present on the surface. \daria{Here we utilize a PHI VersaProbe II surface analysis instrument equipped with an aluminum K-alpha X-ray source. Under the optimal neutralization settings, an overall shift of ${\sim}2.7~\rm{eV}$ from the nominal binding energy values is observed due to the surface charge of the Nb films. The presented data have been adjusted to account for this shift.} We examine the XPS spectrum of $\rm{Nb_{3d}}$ by curve fitting the data using the Lmfit \cite{Newville2014} package. The fits reveal peaks for three distinct niobium oxides (Fig.~\ref{Fig2}(e, f)). \daria{These peaks were fit using the "skewedVoigt" model for asymmetric metallic Nb peaks and the "pseudoVoigt" model for all the other peaks using a Shirley inelastic background with $\rm{3d_{5/2}}$ binding energies located at $202.05$, $203.40$, $205.84$, and $207.38$ eV for Nb, NbO, $\rm{NbO_2}$, and $\rm{Nb_2O_5}$ respectively for the untreated film.} Among those, $\rm{NbO}$ is superconducting, with a transition temperature of $1.38~\rm{K}$ \cite{Hulm1972}. $\rm{Nb_{2}O_{5}}$ is the most thermodynamically stable state of the niobium-oxygen system with the highest binding energy (${\sim}207~\rm{eV}$) and the lowest electrical conductivity \cite{Soares2011}. Due to its various crystalline phases and physical properties, $\rm{Nb_{2}O_5}$ has been considered as one of the main sources of defects present on the surface of Nb \cite{Nico2016}. $\rm{NbO_2}$ also contributes to the loss due to oxygen vacancies\cite{Premkumar2021}, which has the lowest participation in the deposited films reported here.

Prior to placement in the measurement cryostat, devices are cleaned for 7 minutes in an ultrasonic bath of acetone and isopropyl alcohol to remove particles on the samples and strip the protective photoresist. The transfer time to the fridge is kept under 90 minutes for the BOE post-cleaned samples to minimize oxide regrowth on the devices. The samples are placed inside palladium-plated copper microwave launch packaging surrounded by Cryoperm shielding to protect the devices from infrared radiation and external magnetic fields. Mounted devices are cooled inside an adiabatic demagnetization refrigerator (ADR) with a base temperature of $50~\rm{mK}$. 

The resonator transmission, $\rm{S_{21}}$, is measured using a vector network analyzer in an  experimental setup described previously\cite{Zmuidzinas2012,Wisbey2019,Wisbey2010}. Data is collected with the ADR in the temperature regulation mode. Figure \ref{Fig3}(a) displays the transmission near the resonance of a particular device (indicated by the arrow in Fig. \ref{Fig3}(c)) at low power. The quality factors are extracted by employing the $\phi$ rotation method \cite{Gao2008_thesis}. Figure \ref{Fig3}(b) displays the internal loss tangent $\delta_{\rm{int}}$ (inverse of the internal quality factor $\rm{Q_i}$) as a function of average photon number \cite{Sage2011} of the BOE-cleaned device mentioned above. Using the TLS model \cite{Pappas2011}, we extract the filling-factor-adjusted TLS loss tangent ($\rm{F\delta_{TLS}}$), where F is defined as the fraction of the resonator's total loss stored in the TLS material. Figure \ref{Fig3}(c) displays $\rm{F\delta_{TLS}}$ for several devices versus resonance frequency.  The $\rm{F\delta_{TLS}}$ values are an order of magnitude smaller for the case of post-cleaned devices (red circles) than the untreated devices (blue squares) resulting in values as low as $0.15~\rm{ppm}$. The dashed line indicates the average value of $\rm{F\delta_{TLS}}$ of the post-cleaned devices. The average value is $0.32~\rm{ppm}$, well below previous results using the same CPW geometry \cite{McRae2020}. The average low power $\rm{Q_i}$ of the {post-cleaned} devices is $1.13 \times 10^{6}$. In contrast, the untreated devices (blue squares) show an average TLS loss tangent of $1.21~\rm{ppm}$ {with an average single-photon regime $\rm{Q_i}$ of $3.35\times 10^{5}$} comparable with other sputtered niobium studies \cite{Wisbey2010,Burnett_2016}.

Figure \ref{Fig3}(d) displays measurements of the internal loss {of the aforementioned device} versus temperature at fixed power. Based on the TLS model, the TLS-induced loss tends to saturate at high temperatures \cite{Pappas2011,Gao2008} as shown in Fig.~\ref{Fig3}(d). By fitting to the TLS model, we obtain a $\rm{F\delta_{TLS}}$ of $0.44$ ppm in accord with the power scan.

The above results agree with previous findings \cite{Melville2020,Altoe2020,Verjauw2021} that the MA interface is one of the main sources of TLS loss. In our study, we have employed a small CPW gap, which results in a significantly higher concentration of electric field inside the trenches of the CPW, therefore maximizing the coupling with the TLS fluctuators at the interfaces of the material as well as resulting in a larger filling factor value \cite{Sage2011,Chang_2013}. Hence, devices with larger features \cite{Altoe2020} would reduce this filling factor and electric field density thereby reducing $\rm{F\delta_{TLS}}$ by nearly one order of magnitude {as well as yielding higher internal quality factors\cite{Gao2008,Sage2011}}.

We have refined a method to deposit extremely low-loss niobium films for  superconducting CPW resonators using a UHV electron-beam evaporator. With post-cleaning of the Nb surface, devices demonstrated loss tangents well-below previous limits, highlighting the role of the MA interface as one of the main sources of loss in superconducting devices. Moreover, the characterization results show a significant reduction in the surface oxide thickness, verifying the efficacy of our cleaning method. Future work may explore a practical passivation scheme to bypass the MA interface-induced losses for the fabrication of highly coherent superconducting qubit processors.

\begin{acknowledgments}
This research was supported by NSF Grant No. PHY-1752844
(CAREER), NSF Grant No. DMR-1945278 (CAREER), the John
Templeton Foundation Grant No. 61835, and also acknowledge
support from the institute of Materials Science and Engineering at
Washington University.

The data that support the findings of this study are available
from the corresponding author upon reasonable request.
\end{acknowledgments}


\begin{thebibliography}{46}%
\makeatletter
\providecommand \@ifxundefined [1]{%
 \@ifx{#1\undefined}
}%
\providecommand \@ifnum [1]{%
 \ifnum #1\expandafter \@firstoftwo
 \else \expandafter \@secondoftwo
 \fi
}%
\providecommand \@ifx [1]{%
 \ifx #1\expandafter \@firstoftwo
 \else \expandafter \@secondoftwo
 \fi
}%
\providecommand \natexlab [1]{#1}%
\providecommand \enquote  [1]{``#1''}%
\providecommand \bibnamefont  [1]{#1}%
\providecommand \bibfnamefont [1]{#1}%
\providecommand \citenamefont [1]{#1}%
\providecommand \href@noop [0]{\@secondoftwo}%
\providecommand \href [0]{\begingroup \@sanitize@url \@href}%
\providecommand \@href[1]{\@@startlink{#1}\@@href}%
\providecommand \@@href[1]{\endgroup#1\@@endlink}%
\providecommand \@sanitize@url [0]{\catcode `\\12\catcode `\$12\catcode
  `\&12\catcode `\#12\catcode `\^12\catcode `\_12\catcode `\%12\relax}%
\providecommand \@@startlink[1]{}%
\providecommand \@@endlink[0]{}%
\providecommand \url  [0]{\begingroup\@sanitize@url \@url }%
\providecommand \@url [1]{\endgroup\@href {#1}{\urlprefix }}%
\providecommand \urlprefix  [0]{URL }%
\providecommand \Eprint [0]{\href }%
\providecommand \doibase [0]{http://dx.doi.org/}%
\providecommand \selectlanguage [0]{\@gobble}%
\providecommand \bibinfo  [0]{\@secondoftwo}%
\providecommand \bibfield  [0]{\@secondoftwo}%
\providecommand \translation [1]{[#1]}%
\providecommand \BibitemOpen [0]{}%
\providecommand \bibitemStop [0]{}%
\providecommand \bibitemNoStop [0]{.\EOS\space}%
\providecommand \EOS [0]{\spacefactor3000\relax}%
\providecommand \BibitemShut  [1]{\csname bibitem#1\endcsname}%
\let\auto@bib@innerbib\@empty
\bibitem [{\citenamefont {Arute}\ \emph {et~al.}(2019)\citenamefont {Arute},
  \citenamefont {Arya}, \citenamefont {Babbush}, \citenamefont {Bacon},
  \citenamefont {Bardin}, \citenamefont {Barends}, \citenamefont {Biswas},
  \citenamefont {Boixo}, \citenamefont {Brandao}, \citenamefont {Buell},
  \citenamefont {Burkett}, \citenamefont {Chen}, \citenamefont {Chen},
  \citenamefont {Chiaro}, \citenamefont {Collins}, \citenamefont {Courtney},
  \citenamefont {Dunsworth}, \citenamefont {Farhi}, \citenamefont {Foxen},
  \citenamefont {Fowler}, \citenamefont {Gidney}, \citenamefont {Giustina},
  \citenamefont {Graff}, \citenamefont {Guerin}, \citenamefont {Habegger},
  \citenamefont {Harrigan}, \citenamefont {Hartmann}, \citenamefont {Ho},
  \citenamefont {Hoffmann}, \citenamefont {Huang}, \citenamefont {Humble},
  \citenamefont {Isakov}, \citenamefont {Jeffrey}, \citenamefont {Jiang},
  \citenamefont {Kafri}, \citenamefont {Kechedzhi}, \citenamefont {Kelly},
  \citenamefont {Klimov}, \citenamefont {Knysh}, \citenamefont {Korotkov},
  \citenamefont {Kostritsa}, \citenamefont {Landhuis}, \citenamefont
  {Lindmark}, \citenamefont {Lucero}, \citenamefont {Lyakh}, \citenamefont
  {Mandr{\`{a}}}, \citenamefont {McClean}, \citenamefont {McEwen},
  \citenamefont {Megrant}, \citenamefont {Mi}, \citenamefont {Michielsen},
  \citenamefont {Mohseni}, \citenamefont {Mutus}, \citenamefont {Naaman},
  \citenamefont {Neeley}, \citenamefont {Neill}, \citenamefont {Niu},
  \citenamefont {Ostby}, \citenamefont {Petukhov}, \citenamefont {Platt},
  \citenamefont {Quintana}, \citenamefont {Rieffel}, \citenamefont {Roushan},
  \citenamefont {Rubin}, \citenamefont {Sank}, \citenamefont {Satzinger},
  \citenamefont {Smelyanskiy}, \citenamefont {Sung}, \citenamefont
  {Trevithick}, \citenamefont {Vainsencher}, \citenamefont {Villalonga},
  \citenamefont {White}, \citenamefont {Yao}, \citenamefont {Yeh},
  \citenamefont {Zalcman}, \citenamefont {Neven},\ and\ \citenamefont
  {Martinis}}]{Arute2019}%
  \BibitemOpen
  \bibfield  {author} {\bibinfo {author} {\bibfnamefont {F.}~\bibnamefont
  {Arute}}, \bibinfo {author} {\bibfnamefont {K.}~\bibnamefont {Arya}},
  \bibinfo {author} {\bibfnamefont {R.}~\bibnamefont {Babbush}}, \bibinfo
  {author} {\bibfnamefont {D.}~\bibnamefont {Bacon}}, \bibinfo {author}
  {\bibfnamefont {J.~C.}\ \bibnamefont {Bardin}}, \bibinfo {author}
  {\bibfnamefont {R.}~\bibnamefont {Barends}}, \bibinfo {author} {\bibfnamefont
  {R.}~\bibnamefont {Biswas}}, \bibinfo {author} {\bibfnamefont
  {S.}~\bibnamefont {Boixo}}, \bibinfo {author} {\bibfnamefont {F.~G.}\
  \bibnamefont {Brandao}}, \bibinfo {author} {\bibfnamefont {D.~A.}\
  \bibnamefont {Buell}}, \bibinfo {author} {\bibfnamefont {B.}~\bibnamefont
  {Burkett}}, \bibinfo {author} {\bibfnamefont {Y.}~\bibnamefont {Chen}},
  \bibinfo {author} {\bibfnamefont {Z.}~\bibnamefont {Chen}}, \bibinfo {author}
  {\bibfnamefont {B.}~\bibnamefont {Chiaro}}, \bibinfo {author} {\bibfnamefont
  {R.}~\bibnamefont {Collins}}, \bibinfo {author} {\bibfnamefont
  {W.}~\bibnamefont {Courtney}}, \bibinfo {author} {\bibfnamefont
  {A.}~\bibnamefont {Dunsworth}}, \bibinfo {author} {\bibfnamefont
  {E.}~\bibnamefont {Farhi}}, \bibinfo {author} {\bibfnamefont
  {B.}~\bibnamefont {Foxen}}, \bibinfo {author} {\bibfnamefont
  {A.}~\bibnamefont {Fowler}}, \bibinfo {author} {\bibfnamefont
  {C.}~\bibnamefont {Gidney}}, \bibinfo {author} {\bibfnamefont
  {M.}~\bibnamefont {Giustina}}, \bibinfo {author} {\bibfnamefont
  {R.}~\bibnamefont {Graff}}, \bibinfo {author} {\bibfnamefont
  {K.}~\bibnamefont {Guerin}}, \bibinfo {author} {\bibfnamefont
  {S.}~\bibnamefont {Habegger}}, \bibinfo {author} {\bibfnamefont {M.~P.}\
  \bibnamefont {Harrigan}}, \bibinfo {author} {\bibfnamefont {M.~J.}\
  \bibnamefont {Hartmann}}, \bibinfo {author} {\bibfnamefont {A.}~\bibnamefont
  {Ho}}, \bibinfo {author} {\bibfnamefont {M.}~\bibnamefont {Hoffmann}},
  \bibinfo {author} {\bibfnamefont {T.}~\bibnamefont {Huang}}, \bibinfo
  {author} {\bibfnamefont {T.~S.}\ \bibnamefont {Humble}}, \bibinfo {author}
  {\bibfnamefont {S.~V.}\ \bibnamefont {Isakov}}, \bibinfo {author}
  {\bibfnamefont {E.}~\bibnamefont {Jeffrey}}, \bibinfo {author} {\bibfnamefont
  {Z.}~\bibnamefont {Jiang}}, \bibinfo {author} {\bibfnamefont
  {D.}~\bibnamefont {Kafri}}, \bibinfo {author} {\bibfnamefont
  {K.}~\bibnamefont {Kechedzhi}}, \bibinfo {author} {\bibfnamefont
  {J.}~\bibnamefont {Kelly}}, \bibinfo {author} {\bibfnamefont {P.~V.}\
  \bibnamefont {Klimov}}, \bibinfo {author} {\bibfnamefont {S.}~\bibnamefont
  {Knysh}}, \bibinfo {author} {\bibfnamefont {A.}~\bibnamefont {Korotkov}},
  \bibinfo {author} {\bibfnamefont {F.}~\bibnamefont {Kostritsa}}, \bibinfo
  {author} {\bibfnamefont {D.}~\bibnamefont {Landhuis}}, \bibinfo {author}
  {\bibfnamefont {M.}~\bibnamefont {Lindmark}}, \bibinfo {author}
  {\bibfnamefont {E.}~\bibnamefont {Lucero}}, \bibinfo {author} {\bibfnamefont
  {D.}~\bibnamefont {Lyakh}}, \bibinfo {author} {\bibfnamefont
  {S.}~\bibnamefont {Mandr{\`{a}}}}, \bibinfo {author} {\bibfnamefont {J.~R.}\
  \bibnamefont {McClean}}, \bibinfo {author} {\bibfnamefont {M.}~\bibnamefont
  {McEwen}}, \bibinfo {author} {\bibfnamefont {A.}~\bibnamefont {Megrant}},
  \bibinfo {author} {\bibfnamefont {X.}~\bibnamefont {Mi}}, \bibinfo {author}
  {\bibfnamefont {K.}~\bibnamefont {Michielsen}}, \bibinfo {author}
  {\bibfnamefont {M.}~\bibnamefont {Mohseni}}, \bibinfo {author} {\bibfnamefont
  {J.}~\bibnamefont {Mutus}}, \bibinfo {author} {\bibfnamefont
  {O.}~\bibnamefont {Naaman}}, \bibinfo {author} {\bibfnamefont
  {M.}~\bibnamefont {Neeley}}, \bibinfo {author} {\bibfnamefont
  {C.}~\bibnamefont {Neill}}, \bibinfo {author} {\bibfnamefont {M.~Y.}\
  \bibnamefont {Niu}}, \bibinfo {author} {\bibfnamefont {E.}~\bibnamefont
  {Ostby}}, \bibinfo {author} {\bibfnamefont {A.}~\bibnamefont {Petukhov}},
  \bibinfo {author} {\bibfnamefont {J.~C.}\ \bibnamefont {Platt}}, \bibinfo
  {author} {\bibfnamefont {C.}~\bibnamefont {Quintana}}, \bibinfo {author}
  {\bibfnamefont {E.~G.}\ \bibnamefont {Rieffel}}, \bibinfo {author}
  {\bibfnamefont {P.}~\bibnamefont {Roushan}}, \bibinfo {author} {\bibfnamefont
  {N.~C.}\ \bibnamefont {Rubin}}, \bibinfo {author} {\bibfnamefont
  {D.}~\bibnamefont {Sank}}, \bibinfo {author} {\bibfnamefont {K.~J.}\
  \bibnamefont {Satzinger}}, \bibinfo {author} {\bibfnamefont {V.}~\bibnamefont
  {Smelyanskiy}}, \bibinfo {author} {\bibfnamefont {K.~J.}\ \bibnamefont
  {Sung}}, \bibinfo {author} {\bibfnamefont {M.~D.}\ \bibnamefont
  {Trevithick}}, \bibinfo {author} {\bibfnamefont {A.}~\bibnamefont
  {Vainsencher}}, \bibinfo {author} {\bibfnamefont {B.}~\bibnamefont
  {Villalonga}}, \bibinfo {author} {\bibfnamefont {T.}~\bibnamefont {White}},
  \bibinfo {author} {\bibfnamefont {Z.~J.}\ \bibnamefont {Yao}}, \bibinfo
  {author} {\bibfnamefont {P.}~\bibnamefont {Yeh}}, \bibinfo {author}
  {\bibfnamefont {A.}~\bibnamefont {Zalcman}}, \bibinfo {author} {\bibfnamefont
  {H.}~\bibnamefont {Neven}}, \ and\ \bibinfo {author} {\bibfnamefont {J.~M.}\
  \bibnamefont {Martinis}},\ }\bibfield  {title} {\enquote {\bibinfo {title}
  {{Quantum supremacy using a programmable superconducting processor}},}\
  }\href {\doibase 10.1038/s41586-019-1666-5} {\bibfield  {journal} {\bibinfo
  {journal} {Nature}\ }\textbf {\bibinfo {volume} {574}},\ \bibinfo {pages}
  {505--510} (\bibinfo {year} {2019})}\BibitemShut {NoStop}%
\bibitem [{\citenamefont {Jurcevic}\ \emph {et~al.}(2021)\citenamefont
  {Jurcevic}, \citenamefont {Javadi-Abhari}, \citenamefont {Bishop},
  \citenamefont {Lauer}, \citenamefont {Bogorin}, \citenamefont {Brink},
  \citenamefont {Capelluto}, \citenamefont {G{\"{u}}nl{\"{u}}k}, \citenamefont
  {Itoko}, \citenamefont {Kanazawa}, \citenamefont {Kandala}, \citenamefont
  {Keefe}, \citenamefont {Krsulich}, \citenamefont {Landers}, \citenamefont
  {Lewandowski}, \citenamefont {McClure}, \citenamefont {Nannicini},
  \citenamefont {Narasgond}, \citenamefont {Nayfeh}, \citenamefont {Pritchett},
  \citenamefont {Rothwell}, \citenamefont {Srinivasan}, \citenamefont
  {Sundaresan}, \citenamefont {Wang}, \citenamefont {Wei}, \citenamefont
  {Wood}, \citenamefont {Yau}, \citenamefont {Zhang}, \citenamefont {Dial},
  \citenamefont {Chow},\ and\ \citenamefont {Gambetta}}]{Jurcevic2021}%
  \BibitemOpen
  \bibfield  {author} {\bibinfo {author} {\bibfnamefont {P.}~\bibnamefont
  {Jurcevic}}, \bibinfo {author} {\bibfnamefont {A.}~\bibnamefont
  {Javadi-Abhari}}, \bibinfo {author} {\bibfnamefont {L.~S.}\ \bibnamefont
  {Bishop}}, \bibinfo {author} {\bibfnamefont {I.}~\bibnamefont {Lauer}},
  \bibinfo {author} {\bibfnamefont {D.~F.}\ \bibnamefont {Bogorin}}, \bibinfo
  {author} {\bibfnamefont {M.}~\bibnamefont {Brink}}, \bibinfo {author}
  {\bibfnamefont {L.}~\bibnamefont {Capelluto}}, \bibinfo {author}
  {\bibfnamefont {O.}~\bibnamefont {G{\"{u}}nl{\"{u}}k}}, \bibinfo {author}
  {\bibfnamefont {T.}~\bibnamefont {Itoko}}, \bibinfo {author} {\bibfnamefont
  {N.}~\bibnamefont {Kanazawa}}, \bibinfo {author} {\bibfnamefont
  {A.}~\bibnamefont {Kandala}}, \bibinfo {author} {\bibfnamefont {G.~A.}\
  \bibnamefont {Keefe}}, \bibinfo {author} {\bibfnamefont {K.}~\bibnamefont
  {Krsulich}}, \bibinfo {author} {\bibfnamefont {W.}~\bibnamefont {Landers}},
  \bibinfo {author} {\bibfnamefont {E.~P.}\ \bibnamefont {Lewandowski}},
  \bibinfo {author} {\bibfnamefont {D.~T.}\ \bibnamefont {McClure}}, \bibinfo
  {author} {\bibfnamefont {G.}~\bibnamefont {Nannicini}}, \bibinfo {author}
  {\bibfnamefont {A.}~\bibnamefont {Narasgond}}, \bibinfo {author}
  {\bibfnamefont {H.~M.}\ \bibnamefont {Nayfeh}}, \bibinfo {author}
  {\bibfnamefont {E.}~\bibnamefont {Pritchett}}, \bibinfo {author}
  {\bibfnamefont {M.~B.}\ \bibnamefont {Rothwell}}, \bibinfo {author}
  {\bibfnamefont {S.}~\bibnamefont {Srinivasan}}, \bibinfo {author}
  {\bibfnamefont {N.}~\bibnamefont {Sundaresan}}, \bibinfo {author}
  {\bibfnamefont {C.}~\bibnamefont {Wang}}, \bibinfo {author} {\bibfnamefont
  {K.~X.}\ \bibnamefont {Wei}}, \bibinfo {author} {\bibfnamefont {C.~J.}\
  \bibnamefont {Wood}}, \bibinfo {author} {\bibfnamefont {J.~B.}\ \bibnamefont
  {Yau}}, \bibinfo {author} {\bibfnamefont {E.~J.}\ \bibnamefont {Zhang}},
  \bibinfo {author} {\bibfnamefont {O.~E.}\ \bibnamefont {Dial}}, \bibinfo
  {author} {\bibfnamefont {J.~M.}\ \bibnamefont {Chow}}, \ and\ \bibinfo
  {author} {\bibfnamefont {J.~M.}\ \bibnamefont {Gambetta}},\ }\bibfield
  {title} {\enquote {\bibinfo {title} {{Demonstration of quantum volume 64 on a
  superconducting quantum computing system}},}\ }\href {\doibase
  10.1088/2058-9565/abe519} {\bibfield  {journal} {\bibinfo  {journal} {Quantum
  Science and Technology}\ }\textbf {\bibinfo {volume} {6}} (\bibinfo {year}
  {2021}),\ 10.1088/2058-9565/abe519},\ \Eprint
  {http://arxiv.org/abs/2008.08571} {arXiv:2008.08571} \BibitemShut {NoStop}%
\bibitem [{\citenamefont {Wu}\ \emph {et~al.}(2021)\citenamefont {Wu},
  \citenamefont {Bao}, \citenamefont {Cao}, \citenamefont {Chen}, \citenamefont
  {Chen}, \citenamefont {Chen}, \citenamefont {Chung}, \citenamefont {Deng},
  \citenamefont {Du}, \citenamefont {Fan}, \citenamefont {Gong}, \citenamefont
  {Guo}, \citenamefont {Guo}, \citenamefont {Guo}, \citenamefont {Han},
  \citenamefont {Hong}, \citenamefont {Huang}, \citenamefont {Huo},
  \citenamefont {Li}, \citenamefont {Li}, \citenamefont {Li}, \citenamefont
  {Li}, \citenamefont {Liang}, \citenamefont {Lin}, \citenamefont {Lin},
  \citenamefont {Qian}, \citenamefont {Qiao}, \citenamefont {Rong},
  \citenamefont {Su}, \citenamefont {Sun}, \citenamefont {Wang}, \citenamefont
  {Wang}, \citenamefont {Wu}, \citenamefont {Xu}, \citenamefont {Yan},
  \citenamefont {Yang}, \citenamefont {Yang}, \citenamefont {Ye}, \citenamefont
  {Yin}, \citenamefont {Ying}, \citenamefont {Yu}, \citenamefont {Zha},
  \citenamefont {Zhang}, \citenamefont {Zhang}, \citenamefont {Zhang},
  \citenamefont {Zhang}, \citenamefont {Zhao}, \citenamefont {Zhao},
  \citenamefont {Zhou}, \citenamefont {Zhu}, \citenamefont {Lu}, \citenamefont
  {Peng}, \citenamefont {Zhu},\ and\ \citenamefont {Pan}}]{Wu2021}%
  \BibitemOpen
  \bibfield  {author} {\bibinfo {author} {\bibfnamefont {Y.}~\bibnamefont
  {Wu}}, \bibinfo {author} {\bibfnamefont {W.-S.}\ \bibnamefont {Bao}},
  \bibinfo {author} {\bibfnamefont {S.}~\bibnamefont {Cao}}, \bibinfo {author}
  {\bibfnamefont {F.}~\bibnamefont {Chen}}, \bibinfo {author} {\bibfnamefont
  {M.-C.}\ \bibnamefont {Chen}}, \bibinfo {author} {\bibfnamefont
  {X.}~\bibnamefont {Chen}}, \bibinfo {author} {\bibfnamefont {T.-H.}\
  \bibnamefont {Chung}}, \bibinfo {author} {\bibfnamefont {H.}~\bibnamefont
  {Deng}}, \bibinfo {author} {\bibfnamefont {Y.}~\bibnamefont {Du}}, \bibinfo
  {author} {\bibfnamefont {D.}~\bibnamefont {Fan}}, \bibinfo {author}
  {\bibfnamefont {M.}~\bibnamefont {Gong}}, \bibinfo {author} {\bibfnamefont
  {C.}~\bibnamefont {Guo}}, \bibinfo {author} {\bibfnamefont {C.}~\bibnamefont
  {Guo}}, \bibinfo {author} {\bibfnamefont {S.}~\bibnamefont {Guo}}, \bibinfo
  {author} {\bibfnamefont {L.}~\bibnamefont {Han}}, \bibinfo {author}
  {\bibfnamefont {L.}~\bibnamefont {Hong}}, \bibinfo {author} {\bibfnamefont
  {H.-L.}\ \bibnamefont {Huang}}, \bibinfo {author} {\bibfnamefont {Y.-H.}\
  \bibnamefont {Huo}}, \bibinfo {author} {\bibfnamefont {L.}~\bibnamefont
  {Li}}, \bibinfo {author} {\bibfnamefont {N.}~\bibnamefont {Li}}, \bibinfo
  {author} {\bibfnamefont {S.}~\bibnamefont {Li}}, \bibinfo {author}
  {\bibfnamefont {Y.}~\bibnamefont {Li}}, \bibinfo {author} {\bibfnamefont
  {F.}~\bibnamefont {Liang}}, \bibinfo {author} {\bibfnamefont
  {C.}~\bibnamefont {Lin}}, \bibinfo {author} {\bibfnamefont {J.}~\bibnamefont
  {Lin}}, \bibinfo {author} {\bibfnamefont {H.}~\bibnamefont {Qian}}, \bibinfo
  {author} {\bibfnamefont {D.}~\bibnamefont {Qiao}}, \bibinfo {author}
  {\bibfnamefont {H.}~\bibnamefont {Rong}}, \bibinfo {author} {\bibfnamefont
  {H.}~\bibnamefont {Su}}, \bibinfo {author} {\bibfnamefont {L.}~\bibnamefont
  {Sun}}, \bibinfo {author} {\bibfnamefont {L.}~\bibnamefont {Wang}}, \bibinfo
  {author} {\bibfnamefont {S.}~\bibnamefont {Wang}}, \bibinfo {author}
  {\bibfnamefont {D.}~\bibnamefont {Wu}}, \bibinfo {author} {\bibfnamefont
  {Y.}~\bibnamefont {Xu}}, \bibinfo {author} {\bibfnamefont {K.}~\bibnamefont
  {Yan}}, \bibinfo {author} {\bibfnamefont {W.}~\bibnamefont {Yang}}, \bibinfo
  {author} {\bibfnamefont {Y.}~\bibnamefont {Yang}}, \bibinfo {author}
  {\bibfnamefont {Y.}~\bibnamefont {Ye}}, \bibinfo {author} {\bibfnamefont
  {J.}~\bibnamefont {Yin}}, \bibinfo {author} {\bibfnamefont {C.}~\bibnamefont
  {Ying}}, \bibinfo {author} {\bibfnamefont {J.}~\bibnamefont {Yu}}, \bibinfo
  {author} {\bibfnamefont {C.}~\bibnamefont {Zha}}, \bibinfo {author}
  {\bibfnamefont {C.}~\bibnamefont {Zhang}}, \bibinfo {author} {\bibfnamefont
  {H.}~\bibnamefont {Zhang}}, \bibinfo {author} {\bibfnamefont
  {K.}~\bibnamefont {Zhang}}, \bibinfo {author} {\bibfnamefont
  {Y.}~\bibnamefont {Zhang}}, \bibinfo {author} {\bibfnamefont
  {H.}~\bibnamefont {Zhao}}, \bibinfo {author} {\bibfnamefont {Y.}~\bibnamefont
  {Zhao}}, \bibinfo {author} {\bibfnamefont {L.}~\bibnamefont {Zhou}}, \bibinfo
  {author} {\bibfnamefont {Q.}~\bibnamefont {Zhu}}, \bibinfo {author}
  {\bibfnamefont {C.-Y.}\ \bibnamefont {Lu}}, \bibinfo {author} {\bibfnamefont
  {C.-Z.}\ \bibnamefont {Peng}}, \bibinfo {author} {\bibfnamefont
  {X.}~\bibnamefont {Zhu}}, \ and\ \bibinfo {author} {\bibfnamefont {J.-W.}\
  \bibnamefont {Pan}},\ }\bibfield  {title} {\enquote {\bibinfo {title}
  {{Strong quantum computational advantage using a superconducting quantum
  processor}},}\ }\href {http://arxiv.org/abs/2106.14734} {\bibfield  {journal}
  {\bibinfo  {journal} {arXiv preprint}\ } (\bibinfo {year} {2021})},\ \Eprint
  {http://arxiv.org/abs/2106.14734} {arXiv:2106.14734} \BibitemShut {NoStop}%
\bibitem [{\citenamefont {{De Visser}}\ \emph {et~al.}(2014)\citenamefont {{De
  Visser}}, \citenamefont {Goldie}, \citenamefont {Diener}, \citenamefont
  {Withington}, \citenamefont {Baselmans},\ and\ \citenamefont
  {Klapwijk}}]{DeVisser2014}%
  \BibitemOpen
  \bibfield  {author} {\bibinfo {author} {\bibfnamefont {P.~J.}\ \bibnamefont
  {{De Visser}}}, \bibinfo {author} {\bibfnamefont {D.~J.}\ \bibnamefont
  {Goldie}}, \bibinfo {author} {\bibfnamefont {P.}~\bibnamefont {Diener}},
  \bibinfo {author} {\bibfnamefont {S.}~\bibnamefont {Withington}}, \bibinfo
  {author} {\bibfnamefont {J.~J.}\ \bibnamefont {Baselmans}}, \ and\ \bibinfo
  {author} {\bibfnamefont {T.~M.}\ \bibnamefont {Klapwijk}},\ }\bibfield
  {title} {\enquote {\bibinfo {title} {{Evidence of a nonequilibrium
  distribution of quasiparticles in the microwave response of a superconducting
  aluminum resonator}},}\ }\href {\doibase 10.1103/PhysRevLett.112.047004}
  {\bibfield  {journal} {\bibinfo  {journal} {Physical Review Letters}\
  }\textbf {\bibinfo {volume} {112}},\ \bibinfo {pages} {1--5} (\bibinfo {year}
  {2014})},\ \Eprint {http://arxiv.org/abs/1306.4992} {arXiv:1306.4992}
  \BibitemShut {NoStop}%
\bibitem [{\citenamefont {Barends}\ \emph {et~al.}(2011)\citenamefont
  {Barends}, \citenamefont {Wenner}, \citenamefont {Lenander}, \citenamefont
  {Chen}, \citenamefont {Bialczak}, \citenamefont {Kelly}, \citenamefont
  {Lucero}, \citenamefont {O'Malley}, \citenamefont {Mariantoni}, \citenamefont
  {Sank}, \citenamefont {Wang}, \citenamefont {White}, \citenamefont {Yin},
  \citenamefont {Zhao}, \citenamefont {Cleland}, \citenamefont {Martinis},\
  and\ \citenamefont {Baselmans}}]{Barends2011}%
  \BibitemOpen
  \bibfield  {author} {\bibinfo {author} {\bibfnamefont {R.}~\bibnamefont
  {Barends}}, \bibinfo {author} {\bibfnamefont {J.}~\bibnamefont {Wenner}},
  \bibinfo {author} {\bibfnamefont {M.}~\bibnamefont {Lenander}}, \bibinfo
  {author} {\bibfnamefont {Y.}~\bibnamefont {Chen}}, \bibinfo {author}
  {\bibfnamefont {R.~C.}\ \bibnamefont {Bialczak}}, \bibinfo {author}
  {\bibfnamefont {J.}~\bibnamefont {Kelly}}, \bibinfo {author} {\bibfnamefont
  {E.}~\bibnamefont {Lucero}}, \bibinfo {author} {\bibfnamefont
  {P.}~\bibnamefont {O'Malley}}, \bibinfo {author} {\bibfnamefont
  {M.}~\bibnamefont {Mariantoni}}, \bibinfo {author} {\bibfnamefont
  {D.}~\bibnamefont {Sank}}, \bibinfo {author} {\bibfnamefont {H.}~\bibnamefont
  {Wang}}, \bibinfo {author} {\bibfnamefont {T.~C.}\ \bibnamefont {White}},
  \bibinfo {author} {\bibfnamefont {Y.}~\bibnamefont {Yin}}, \bibinfo {author}
  {\bibfnamefont {J.}~\bibnamefont {Zhao}}, \bibinfo {author} {\bibfnamefont
  {A.~N.}\ \bibnamefont {Cleland}}, \bibinfo {author} {\bibfnamefont {J.~M.}\
  \bibnamefont {Martinis}}, \ and\ \bibinfo {author} {\bibfnamefont {J.~J.}\
  \bibnamefont {Baselmans}},\ }\bibfield  {title} {\enquote {\bibinfo {title}
  {{Minimizing quasiparticle generation from stray infrared light in
  superconducting quantum circuits}},}\ }\href {\doibase 10.1063/1.3638063}
  {\bibfield  {journal} {\bibinfo  {journal} {Applied Physics Letters}\
  }\textbf {\bibinfo {volume} {99}},\ \bibinfo {pages} {1--4} (\bibinfo {year}
  {2011})}\BibitemShut {NoStop}%
\bibitem [{\citenamefont {C{\'{o}}rcoles}\ \emph {et~al.}(2011)\citenamefont
  {C{\'{o}}rcoles}, \citenamefont {Chow}, \citenamefont {Gambetta},
  \citenamefont {Rigetti}, \citenamefont {Rozen}, \citenamefont {Keefe},
  \citenamefont {{Beth Rothwell}}, \citenamefont {Ketchen},\ and\ \citenamefont
  {Steffen}}]{Corcoles2011}%
  \BibitemOpen
  \bibfield  {author} {\bibinfo {author} {\bibfnamefont {A.~D.}\ \bibnamefont
  {C{\'{o}}rcoles}}, \bibinfo {author} {\bibfnamefont {J.~M.}\ \bibnamefont
  {Chow}}, \bibinfo {author} {\bibfnamefont {J.~M.}\ \bibnamefont {Gambetta}},
  \bibinfo {author} {\bibfnamefont {C.}~\bibnamefont {Rigetti}}, \bibinfo
  {author} {\bibfnamefont {J.~R.}\ \bibnamefont {Rozen}}, \bibinfo {author}
  {\bibfnamefont {G.~A.}\ \bibnamefont {Keefe}}, \bibinfo {author}
  {\bibfnamefont {M.}~\bibnamefont {{Beth Rothwell}}}, \bibinfo {author}
  {\bibfnamefont {M.~B.}\ \bibnamefont {Ketchen}}, \ and\ \bibinfo {author}
  {\bibfnamefont {M.}~\bibnamefont {Steffen}},\ }\bibfield  {title} {\enquote
  {\bibinfo {title} {{Protecting superconducting qubits from radiation}},}\
  }\href {\doibase 10.1063/1.3658630} {\bibfield  {journal} {\bibinfo
  {journal} {Applied Physics Letters}\ }\textbf {\bibinfo {volume} {99}}
  (\bibinfo {year} {2011}),\ 10.1063/1.3658630}\BibitemShut {NoStop}%
\bibitem [{\citenamefont {Song}\ \emph {et~al.}(2009)\citenamefont {Song},
  \citenamefont {Heitmann}, \citenamefont {Defeo}, \citenamefont {Yu},
  \citenamefont {McDermott}, \citenamefont {Neeley}, \citenamefont {Martinis},\
  and\ \citenamefont {Plourde}}]{Song2009}%
  \BibitemOpen
  \bibfield  {author} {\bibinfo {author} {\bibfnamefont {C.}~\bibnamefont
  {Song}}, \bibinfo {author} {\bibfnamefont {T.~W.}\ \bibnamefont {Heitmann}},
  \bibinfo {author} {\bibfnamefont {M.~P.}\ \bibnamefont {Defeo}}, \bibinfo
  {author} {\bibfnamefont {K.}~\bibnamefont {Yu}}, \bibinfo {author}
  {\bibfnamefont {R.}~\bibnamefont {McDermott}}, \bibinfo {author}
  {\bibfnamefont {M.}~\bibnamefont {Neeley}}, \bibinfo {author} {\bibfnamefont
  {J.~M.}\ \bibnamefont {Martinis}}, \ and\ \bibinfo {author} {\bibfnamefont
  {B.~L.}\ \bibnamefont {Plourde}},\ }\bibfield  {title} {\enquote {\bibinfo
  {title} {{Microwave response of vortices in superconducting thin films of Re
  and Al}},}\ }\href {\doibase 10.1103/PhysRevB.79.174512} {\bibfield
  {journal} {\bibinfo  {journal} {Physical Review B - Condensed Matter and
  Materials Physics}\ }\textbf {\bibinfo {volume} {79}},\ \bibinfo {pages}
  {1--9} (\bibinfo {year} {2009})},\ \Eprint {http://arxiv.org/abs/0812.3645}
  {arXiv:0812.3645} \BibitemShut {NoStop}%
\bibitem [{\citenamefont {Sage}\ \emph {et~al.}(2011)\citenamefont {Sage},
  \citenamefont {Bolkhovsky}, \citenamefont {Oliver}, \citenamefont {Turek},\
  and\ \citenamefont {Welander}}]{Sage2011}%
  \BibitemOpen
  \bibfield  {author} {\bibinfo {author} {\bibfnamefont {J.~M.}\ \bibnamefont
  {Sage}}, \bibinfo {author} {\bibfnamefont {V.}~\bibnamefont {Bolkhovsky}},
  \bibinfo {author} {\bibfnamefont {W.~D.}\ \bibnamefont {Oliver}}, \bibinfo
  {author} {\bibfnamefont {B.}~\bibnamefont {Turek}}, \ and\ \bibinfo {author}
  {\bibfnamefont {P.~B.}\ \bibnamefont {Welander}},\ }\bibfield  {title}
  {\enquote {\bibinfo {title} {{Study of loss in superconducting coplanar
  waveguide resonators}},}\ }\href {\doibase 10.1063/1.3552890} {\bibfield
  {journal} {\bibinfo  {journal} {Journal of Applied Physics}\ } (\bibinfo
  {year} {2011}),\ 10.1063/1.3552890},\ \Eprint
  {http://arxiv.org/abs/1010.6063} {arXiv:1010.6063} \BibitemShut {NoStop}%
\bibitem [{\citenamefont {Kreikebaum}\ \emph {et~al.}(2016)\citenamefont
  {Kreikebaum}, \citenamefont {Dove}, \citenamefont {Livingston}, \citenamefont
  {Kim},\ and\ \citenamefont {Siddiqi}}]{Kreikebaum2016}%
  \BibitemOpen
  \bibfield  {author} {\bibinfo {author} {\bibfnamefont {J.~M.}\ \bibnamefont
  {Kreikebaum}}, \bibinfo {author} {\bibfnamefont {A.}~\bibnamefont {Dove}},
  \bibinfo {author} {\bibfnamefont {W.}~\bibnamefont {Livingston}}, \bibinfo
  {author} {\bibfnamefont {E.}~\bibnamefont {Kim}}, \ and\ \bibinfo {author}
  {\bibfnamefont {I.}~\bibnamefont {Siddiqi}},\ }\bibfield  {title} {\enquote
  {\bibinfo {title} {{Optimization of infrared and magnetic shielding of
  superconducting TiN and Al coplanar microwave resonators}},}\ }\href
  {\doibase 10.1088/0953-2048/29/10/104002} {\bibfield  {journal} {\bibinfo
  {journal} {Superconductor Science and Technology}\ }\textbf {\bibinfo
  {volume} {29}} (\bibinfo {year} {2016}),\
  10.1088/0953-2048/29/10/104002}\BibitemShut {NoStop}%
\bibitem [{\citenamefont {Sandberg}\ \emph {et~al.}(2013)\citenamefont
  {Sandberg}, \citenamefont {Vissers}, \citenamefont {Ohki}, \citenamefont
  {Gao}, \citenamefont {Aumentado}, \citenamefont {Weides},\ and\ \citenamefont
  {Pappas}}]{Sandberg2013}%
  \BibitemOpen
  \bibfield  {author} {\bibinfo {author} {\bibfnamefont {M.}~\bibnamefont
  {Sandberg}}, \bibinfo {author} {\bibfnamefont {M.~R.}\ \bibnamefont
  {Vissers}}, \bibinfo {author} {\bibfnamefont {T.~A.}\ \bibnamefont {Ohki}},
  \bibinfo {author} {\bibfnamefont {J.}~\bibnamefont {Gao}}, \bibinfo {author}
  {\bibfnamefont {J.}~\bibnamefont {Aumentado}}, \bibinfo {author}
  {\bibfnamefont {M.}~\bibnamefont {Weides}}, \ and\ \bibinfo {author}
  {\bibfnamefont {D.~P.}\ \bibnamefont {Pappas}},\ }\bibfield  {title}
  {\enquote {\bibinfo {title} {{Radiation-suppressed superconducting quantum
  bit in a planar geometry}},}\ }\href {\doibase 10.1063/1.4792698} {\bibfield
  {journal} {\bibinfo  {journal} {Applied Physics Letters}\ }\textbf {\bibinfo
  {volume} {102}} (\bibinfo {year} {2013}),\ 10.1063/1.4792698}\BibitemShut
  {NoStop}%
\bibitem [{\citenamefont {Chiaro}\ \emph {et~al.}(2016)\citenamefont {Chiaro},
  \citenamefont {Megrant}, \citenamefont {Dunsworth}, \citenamefont {Chen},
  \citenamefont {Barends}, \citenamefont {Campbell}, \citenamefont {Chen},
  \citenamefont {Fowler}, \citenamefont {Hoi}, \citenamefont {Jeffrey},
  \citenamefont {Kelly}, \citenamefont {Mutus}, \citenamefont {Neill},
  \citenamefont {O'Malley}, \citenamefont {Quintana}, \citenamefont {Roushan},
  \citenamefont {Sank}, \citenamefont {Vainsencher}, \citenamefont {Wenner},
  \citenamefont {White},\ and\ \citenamefont {Martinis}}]{Chiaro2016}%
  \BibitemOpen
  \bibfield  {author} {\bibinfo {author} {\bibfnamefont {B.}~\bibnamefont
  {Chiaro}}, \bibinfo {author} {\bibfnamefont {A.}~\bibnamefont {Megrant}},
  \bibinfo {author} {\bibfnamefont {A.}~\bibnamefont {Dunsworth}}, \bibinfo
  {author} {\bibfnamefont {Z.}~\bibnamefont {Chen}}, \bibinfo {author}
  {\bibfnamefont {R.}~\bibnamefont {Barends}}, \bibinfo {author} {\bibfnamefont
  {B.}~\bibnamefont {Campbell}}, \bibinfo {author} {\bibfnamefont
  {Y.}~\bibnamefont {Chen}}, \bibinfo {author} {\bibfnamefont {A.}~\bibnamefont
  {Fowler}}, \bibinfo {author} {\bibfnamefont {I.~C.}\ \bibnamefont {Hoi}},
  \bibinfo {author} {\bibfnamefont {E.}~\bibnamefont {Jeffrey}}, \bibinfo
  {author} {\bibfnamefont {J.}~\bibnamefont {Kelly}}, \bibinfo {author}
  {\bibfnamefont {J.}~\bibnamefont {Mutus}}, \bibinfo {author} {\bibfnamefont
  {C.}~\bibnamefont {Neill}}, \bibinfo {author} {\bibfnamefont {P.~J.}\
  \bibnamefont {O'Malley}}, \bibinfo {author} {\bibfnamefont {C.}~\bibnamefont
  {Quintana}}, \bibinfo {author} {\bibfnamefont {P.}~\bibnamefont {Roushan}},
  \bibinfo {author} {\bibfnamefont {D.}~\bibnamefont {Sank}}, \bibinfo {author}
  {\bibfnamefont {A.}~\bibnamefont {Vainsencher}}, \bibinfo {author}
  {\bibfnamefont {J.}~\bibnamefont {Wenner}}, \bibinfo {author} {\bibfnamefont
  {T.~C.}\ \bibnamefont {White}}, \ and\ \bibinfo {author} {\bibfnamefont
  {J.~M.}\ \bibnamefont {Martinis}},\ }\bibfield  {title} {\enquote {\bibinfo
  {title} {{Dielectric surface loss in superconducting resonators with
  flux-trapping holes}},}\ }\href {\doibase 10.1088/0953-2048/29/10/104006}
  {\bibfield  {journal} {\bibinfo  {journal} {Superconductor Science and
  Technology}\ }\textbf {\bibinfo {volume} {29}} (\bibinfo {year} {2016}),\
  10.1088/0953-2048/29/10/104006},\ \Eprint {http://arxiv.org/abs/1607.05841}
  {arXiv:1607.05841} \BibitemShut {NoStop}%
\bibitem [{\citenamefont {Lisenfeld}\ \emph {et~al.}(2019)\citenamefont
  {Lisenfeld}, \citenamefont {Bilmes}, \citenamefont {Megrant}, \citenamefont
  {Barends}, \citenamefont {Kelly}, \citenamefont {Klimov}, \citenamefont
  {Weiss}, \citenamefont {Martinis},\ and\ \citenamefont
  {Ustinov}}]{Lisenfeld2019}%
  \BibitemOpen
  \bibfield  {author} {\bibinfo {author} {\bibfnamefont {J.}~\bibnamefont
  {Lisenfeld}}, \bibinfo {author} {\bibfnamefont {A.}~\bibnamefont {Bilmes}},
  \bibinfo {author} {\bibfnamefont {A.}~\bibnamefont {Megrant}}, \bibinfo
  {author} {\bibfnamefont {R.}~\bibnamefont {Barends}}, \bibinfo {author}
  {\bibfnamefont {J.}~\bibnamefont {Kelly}}, \bibinfo {author} {\bibfnamefont
  {P.}~\bibnamefont {Klimov}}, \bibinfo {author} {\bibfnamefont
  {G.}~\bibnamefont {Weiss}}, \bibinfo {author} {\bibfnamefont {J.~M.}\
  \bibnamefont {Martinis}}, \ and\ \bibinfo {author} {\bibfnamefont {A.~V.}\
  \bibnamefont {Ustinov}},\ }\bibfield  {title} {\enquote {\bibinfo {title}
  {{Electric field spectroscopy of material defects in transmon qubits}},}\
  }\href {\doibase 10.1038/s41534-019-0224-1} {\bibfield  {journal} {\bibinfo
  {journal} {npj Quantum Information}\ }\textbf {\bibinfo {volume} {5}},\
  \bibinfo {pages} {1--6} (\bibinfo {year} {2019})},\ \Eprint
  {http://arxiv.org/abs/1909.09749} {arXiv:1909.09749} \BibitemShut {NoStop}%
\bibitem [{\citenamefont {de~Graaf}\ \emph {et~al.}(2020)\citenamefont
  {de~Graaf}, \citenamefont {Faoro}, \citenamefont {Ioffe}, \citenamefont
  {Mahashabde}, \citenamefont {Burnett}, \citenamefont {Lindstr{\"{o}}m},
  \citenamefont {Kubatkin}, \citenamefont {Danilov},\ and\ \citenamefont
  {Tzalenchuk}}]{DeGraaf2020}%
  \BibitemOpen
  \bibfield  {author} {\bibinfo {author} {\bibfnamefont {S.~E.}\ \bibnamefont
  {de~Graaf}}, \bibinfo {author} {\bibfnamefont {L.}~\bibnamefont {Faoro}},
  \bibinfo {author} {\bibfnamefont {L.~B.}\ \bibnamefont {Ioffe}}, \bibinfo
  {author} {\bibfnamefont {S.}~\bibnamefont {Mahashabde}}, \bibinfo {author}
  {\bibfnamefont {J.~J.}\ \bibnamefont {Burnett}}, \bibinfo {author}
  {\bibfnamefont {T.}~\bibnamefont {Lindstr{\"{o}}m}}, \bibinfo {author}
  {\bibfnamefont {S.~E.}\ \bibnamefont {Kubatkin}}, \bibinfo {author}
  {\bibfnamefont {A.~V.}\ \bibnamefont {Danilov}}, \ and\ \bibinfo {author}
  {\bibfnamefont {A.~Y.}\ \bibnamefont {Tzalenchuk}},\ }\bibfield  {title}
  {\enquote {\bibinfo {title} {{Two-level systems in superconducting quantum
  devices due to trapped quasiparticles}},}\ }\href {\doibase
  10.1126/SCIADV.ABC5055} {\bibfield  {journal} {\bibinfo  {journal} {Science
  Advances}\ }\textbf {\bibinfo {volume} {6}},\ \bibinfo {pages} {1--9}
  (\bibinfo {year} {2020})},\ \Eprint {http://arxiv.org/abs/2004.02485}
  {arXiv:2004.02485} \BibitemShut {NoStop}%
\bibitem [{\citenamefont {Niepce}\ \emph {et~al.}(2020)\citenamefont {Niepce},
  \citenamefont {Burnett}, \citenamefont {Latorre},\ and\ \citenamefont
  {Bylander}}]{Niepce2020}%
  \BibitemOpen
  \bibfield  {author} {\bibinfo {author} {\bibfnamefont {D.}~\bibnamefont
  {Niepce}}, \bibinfo {author} {\bibfnamefont {J.~J.}\ \bibnamefont {Burnett}},
  \bibinfo {author} {\bibfnamefont {M.~G.}\ \bibnamefont {Latorre}}, \ and\
  \bibinfo {author} {\bibfnamefont {J.}~\bibnamefont {Bylander}},\ }\bibfield
  {title} {\enquote {\bibinfo {title} {{Geometric scaling of two-level-system
  loss in superconducting resonators}},}\ }\href {\doibase
  10.1088/1361-6668/ab6179} {\bibfield  {journal} {\bibinfo  {journal}
  {Superconductor Science and Technology}\ }\textbf {\bibinfo {volume} {33}}
  (\bibinfo {year} {2020}),\ 10.1088/1361-6668/ab6179},\ \Eprint
  {http://arxiv.org/abs/1908.02606} {arXiv:1908.02606} \BibitemShut {NoStop}%
\bibitem [{\citenamefont {Gao}\ \emph {et~al.}(2008)\citenamefont {Gao},
  \citenamefont {Daal}, \citenamefont {Vayonakis}, \citenamefont {Kumar},
  \citenamefont {Zmuidzinas}, \citenamefont {Sadoulet}, \citenamefont {Mazin},
  \citenamefont {Day},\ and\ \citenamefont {Leduc}}]{Gao2008}%
  \BibitemOpen
  \bibfield  {author} {\bibinfo {author} {\bibfnamefont {J.}~\bibnamefont
  {Gao}}, \bibinfo {author} {\bibfnamefont {M.}~\bibnamefont {Daal}}, \bibinfo
  {author} {\bibfnamefont {A.}~\bibnamefont {Vayonakis}}, \bibinfo {author}
  {\bibfnamefont {S.}~\bibnamefont {Kumar}}, \bibinfo {author} {\bibfnamefont
  {J.}~\bibnamefont {Zmuidzinas}}, \bibinfo {author} {\bibfnamefont
  {B.}~\bibnamefont {Sadoulet}}, \bibinfo {author} {\bibfnamefont {B.~A.}\
  \bibnamefont {Mazin}}, \bibinfo {author} {\bibfnamefont {P.~K.}\ \bibnamefont
  {Day}}, \ and\ \bibinfo {author} {\bibfnamefont {H.~G.}\ \bibnamefont
  {Leduc}},\ }\bibfield  {title} {\enquote {\bibinfo {title} {{Experimental
  evidence for a surface distribution of two-level systems in superconducting
  lithographed microwave resonators}},}\ }\href {\doibase 10.1063/1.2906373}
  {\bibfield  {journal} {\bibinfo  {journal} {Applied Physics Letters}\
  }\textbf {\bibinfo {volume} {92}} (\bibinfo {year} {2008}),\
  10.1063/1.2906373},\ \Eprint {http://arxiv.org/abs/0802.4457}
  {arXiv:0802.4457} \BibitemShut {NoStop}%
\bibitem [{\citenamefont {Wisbey}\ \emph {et~al.}(2010)\citenamefont {Wisbey},
  \citenamefont {Gao}, \citenamefont {Vissers}, \citenamefont {{Da Silva}},
  \citenamefont {Kline}, \citenamefont {Vale},\ and\ \citenamefont
  {Pappas}}]{Wisbey2010}%
  \BibitemOpen
  \bibfield  {author} {\bibinfo {author} {\bibfnamefont {D.~S.}\ \bibnamefont
  {Wisbey}}, \bibinfo {author} {\bibfnamefont {J.}~\bibnamefont {Gao}},
  \bibinfo {author} {\bibfnamefont {M.~R.}\ \bibnamefont {Vissers}}, \bibinfo
  {author} {\bibfnamefont {F.~C.}\ \bibnamefont {{Da Silva}}}, \bibinfo
  {author} {\bibfnamefont {J.~S.}\ \bibnamefont {Kline}}, \bibinfo {author}
  {\bibfnamefont {L.}~\bibnamefont {Vale}}, \ and\ \bibinfo {author}
  {\bibfnamefont {D.~P.}\ \bibnamefont {Pappas}},\ }\bibfield  {title}
  {\enquote {\bibinfo {title} {{Effect of metal/substrate interfaces on
  radio-frequency loss in superconducting coplanar waveguides}},}\ }\href
  {\doibase 10.1063/1.3499608} {\bibfield  {journal} {\bibinfo  {journal}
  {Journal of Applied Physics}\ }\textbf {\bibinfo {volume} {108}},\ \bibinfo
  {pages} {8--12} (\bibinfo {year} {2010})}\BibitemShut {NoStop}%
\bibitem [{\citenamefont {Bilmes}\ \emph {et~al.}(2020)\citenamefont {Bilmes},
  \citenamefont {Megrant}, \citenamefont {Klimov}, \citenamefont {Weiss},
  \citenamefont {Martinis}, \citenamefont {Ustinov},\ and\ \citenamefont
  {Lisenfeld}}]{Bilmes2020}%
  \BibitemOpen
  \bibfield  {author} {\bibinfo {author} {\bibfnamefont {A.}~\bibnamefont
  {Bilmes}}, \bibinfo {author} {\bibfnamefont {A.}~\bibnamefont {Megrant}},
  \bibinfo {author} {\bibfnamefont {P.}~\bibnamefont {Klimov}}, \bibinfo
  {author} {\bibfnamefont {G.}~\bibnamefont {Weiss}}, \bibinfo {author}
  {\bibfnamefont {J.~M.}\ \bibnamefont {Martinis}}, \bibinfo {author}
  {\bibfnamefont {A.~V.}\ \bibnamefont {Ustinov}}, \ and\ \bibinfo {author}
  {\bibfnamefont {J.}~\bibnamefont {Lisenfeld}},\ }\bibfield  {title} {\enquote
  {\bibinfo {title} {{Resolving the positions of defects in superconducting
  quantum bits}},}\ }\href {\doibase 10.1038/s41598-020-59749-y} {\bibfield
  {journal} {\bibinfo  {journal} {Scientific Reports}\ }\textbf {\bibinfo
  {volume} {10}},\ \bibinfo {pages} {1--6} (\bibinfo {year} {2020})},\ \Eprint
  {http://arxiv.org/abs/1911.08246} {arXiv:1911.08246} \BibitemShut {NoStop}%
\bibitem [{\citenamefont {Gr{\"{u}}nhaupt}\ \emph {et~al.}(2017)\citenamefont
  {Gr{\"{u}}nhaupt}, \citenamefont {{Von L{\"{u}}pke}}, \citenamefont
  {Gusenkova}, \citenamefont {Skacel}, \citenamefont {Maleeva}, \citenamefont
  {Schl{\"{o}}r}, \citenamefont {Bilmes}, \citenamefont {Rotzinger},
  \citenamefont {Ustinov}, \citenamefont {Weides},\ and\ \citenamefont
  {Pop}}]{Grunhaupt2017}%
  \BibitemOpen
  \bibfield  {author} {\bibinfo {author} {\bibfnamefont {L.}~\bibnamefont
  {Gr{\"{u}}nhaupt}}, \bibinfo {author} {\bibfnamefont {U.}~\bibnamefont {{Von
  L{\"{u}}pke}}}, \bibinfo {author} {\bibfnamefont {D.}~\bibnamefont
  {Gusenkova}}, \bibinfo {author} {\bibfnamefont {S.~T.}\ \bibnamefont
  {Skacel}}, \bibinfo {author} {\bibfnamefont {N.}~\bibnamefont {Maleeva}},
  \bibinfo {author} {\bibfnamefont {S.}~\bibnamefont {Schl{\"{o}}r}}, \bibinfo
  {author} {\bibfnamefont {A.}~\bibnamefont {Bilmes}}, \bibinfo {author}
  {\bibfnamefont {H.}~\bibnamefont {Rotzinger}}, \bibinfo {author}
  {\bibfnamefont {A.~V.}\ \bibnamefont {Ustinov}}, \bibinfo {author}
  {\bibfnamefont {M.}~\bibnamefont {Weides}}, \ and\ \bibinfo {author}
  {\bibfnamefont {I.~M.}\ \bibnamefont {Pop}},\ }\bibfield  {title} {\enquote
  {\bibinfo {title} {{An argon ion beam milling process for native AlOx layers
  enabling coherent superconducting contacts}},}\ }\href {\doibase
  10.1063/1.4990491} {\bibfield  {journal} {\bibinfo  {journal} {Applied
  Physics Letters}\ }\textbf {\bibinfo {volume} {111}} (\bibinfo {year}
  {2017}),\ 10.1063/1.4990491},\ \Eprint {http://arxiv.org/abs/1706.06424}
  {arXiv:1706.06424} \BibitemShut {NoStop}%
\bibitem [{\citenamefont {Calusine}\ \emph {et~al.}(2018)\citenamefont
  {Calusine}, \citenamefont {Melville}, \citenamefont {Woods}, \citenamefont
  {Das}, \citenamefont {Stull}, \citenamefont {Bolkhovsky}, \citenamefont
  {Braje}, \citenamefont {Hover}, \citenamefont {Kim}, \citenamefont {Miloshi},
  \citenamefont {Rosenberg}, \citenamefont {Sevi}, \citenamefont {Yoder},
  \citenamefont {Dauler},\ and\ \citenamefont {Oliver}}]{Calusine2018}%
  \BibitemOpen
  \bibfield  {author} {\bibinfo {author} {\bibfnamefont {G.}~\bibnamefont
  {Calusine}}, \bibinfo {author} {\bibfnamefont {A.}~\bibnamefont {Melville}},
  \bibinfo {author} {\bibfnamefont {W.}~\bibnamefont {Woods}}, \bibinfo
  {author} {\bibfnamefont {R.}~\bibnamefont {Das}}, \bibinfo {author}
  {\bibfnamefont {C.}~\bibnamefont {Stull}}, \bibinfo {author} {\bibfnamefont
  {V.}~\bibnamefont {Bolkhovsky}}, \bibinfo {author} {\bibfnamefont
  {D.}~\bibnamefont {Braje}}, \bibinfo {author} {\bibfnamefont
  {D.}~\bibnamefont {Hover}}, \bibinfo {author} {\bibfnamefont {D.~K.}\
  \bibnamefont {Kim}}, \bibinfo {author} {\bibfnamefont {X.}~\bibnamefont
  {Miloshi}}, \bibinfo {author} {\bibfnamefont {D.}~\bibnamefont {Rosenberg}},
  \bibinfo {author} {\bibfnamefont {A.}~\bibnamefont {Sevi}}, \bibinfo {author}
  {\bibfnamefont {J.~L.}\ \bibnamefont {Yoder}}, \bibinfo {author}
  {\bibfnamefont {E.}~\bibnamefont {Dauler}}, \ and\ \bibinfo {author}
  {\bibfnamefont {W.~D.}\ \bibnamefont {Oliver}},\ }\bibfield  {title}
  {\enquote {\bibinfo {title} {{Analysis and mitigation of interface losses in
  trenched superconducting coplanar waveguide resonators}},}\ }\href {\doibase
  10.1063/1.5006888} {\bibfield  {journal} {\bibinfo  {journal} {Applied
  Physics Letters}\ }\textbf {\bibinfo {volume} {112}} (\bibinfo {year}
  {2018}),\ 10.1063/1.5006888},\ \Eprint {http://arxiv.org/abs/1709.10015}
  {arXiv:1709.10015} \BibitemShut {NoStop}%
\bibitem [{\citenamefont {Place}\ \emph {et~al.}(2021)\citenamefont {Place},
  \citenamefont {Rodgers}, \citenamefont {Mundada}, \citenamefont {Smitham},
  \citenamefont {Fitzpatrick}, \citenamefont {Leng}, \citenamefont {Premkumar},
  \citenamefont {Bryon}, \citenamefont {Vrajitoarea}, \citenamefont {Sussman},
  \citenamefont {Cheng}, \citenamefont {Madhavan}, \citenamefont {Babla},
  \citenamefont {Le}, \citenamefont {Gang}, \citenamefont {J{\"{a}}ck},
  \citenamefont {Gyenis}, \citenamefont {Yao}, \citenamefont {Cava},
  \citenamefont {de~Leon},\ and\ \citenamefont {Houck}}]{Place2021}%
  \BibitemOpen
  \bibfield  {author} {\bibinfo {author} {\bibfnamefont {A.~P.}\ \bibnamefont
  {Place}}, \bibinfo {author} {\bibfnamefont {L.~V.}\ \bibnamefont {Rodgers}},
  \bibinfo {author} {\bibfnamefont {P.}~\bibnamefont {Mundada}}, \bibinfo
  {author} {\bibfnamefont {B.~M.}\ \bibnamefont {Smitham}}, \bibinfo {author}
  {\bibfnamefont {M.}~\bibnamefont {Fitzpatrick}}, \bibinfo {author}
  {\bibfnamefont {Z.}~\bibnamefont {Leng}}, \bibinfo {author} {\bibfnamefont
  {A.}~\bibnamefont {Premkumar}}, \bibinfo {author} {\bibfnamefont
  {J.}~\bibnamefont {Bryon}}, \bibinfo {author} {\bibfnamefont
  {A.}~\bibnamefont {Vrajitoarea}}, \bibinfo {author} {\bibfnamefont
  {S.}~\bibnamefont {Sussman}}, \bibinfo {author} {\bibfnamefont
  {G.}~\bibnamefont {Cheng}}, \bibinfo {author} {\bibfnamefont
  {T.}~\bibnamefont {Madhavan}}, \bibinfo {author} {\bibfnamefont {H.~K.}\
  \bibnamefont {Babla}}, \bibinfo {author} {\bibfnamefont {X.~H.}\ \bibnamefont
  {Le}}, \bibinfo {author} {\bibfnamefont {Y.}~\bibnamefont {Gang}}, \bibinfo
  {author} {\bibfnamefont {B.}~\bibnamefont {J{\"{a}}ck}}, \bibinfo {author}
  {\bibfnamefont {A.}~\bibnamefont {Gyenis}}, \bibinfo {author} {\bibfnamefont
  {N.}~\bibnamefont {Yao}}, \bibinfo {author} {\bibfnamefont {R.~J.}\
  \bibnamefont {Cava}}, \bibinfo {author} {\bibfnamefont {N.~P.}\ \bibnamefont
  {de~Leon}}, \ and\ \bibinfo {author} {\bibfnamefont {A.~A.}\ \bibnamefont
  {Houck}},\ }\bibfield  {title} {\enquote {\bibinfo {title} {{New material
  platform for superconducting transmon qubits with coherence times exceeding
  0.3 milliseconds}},}\ }\href {\doibase 10.1038/s41467-021-22030-5} {\bibfield
   {journal} {\bibinfo  {journal} {Nature Communications}\ }\textbf {\bibinfo
  {volume} {12}} (\bibinfo {year} {2021}),\ 10.1038/s41467-021-22030-5},\
  \Eprint {http://arxiv.org/abs/2003.00024} {arXiv:2003.00024} \BibitemShut
  {NoStop}%
\bibitem [{\citenamefont {Monroe}\ \emph {et~al.}(2021)\citenamefont {Monroe},
  \citenamefont {Kowsari}, \citenamefont {Zheng}, \citenamefont {Gaikwad},
  \citenamefont {Brewster}, \citenamefont {Wisbey},\ and\ \citenamefont
  {Murch}}]{Monroe2021}%
  \BibitemOpen
  \bibfield  {author} {\bibinfo {author} {\bibfnamefont {J.~T.}\ \bibnamefont
  {Monroe}}, \bibinfo {author} {\bibfnamefont {D.}~\bibnamefont {Kowsari}},
  \bibinfo {author} {\bibfnamefont {K.}~\bibnamefont {Zheng}}, \bibinfo
  {author} {\bibfnamefont {C.}~\bibnamefont {Gaikwad}}, \bibinfo {author}
  {\bibfnamefont {J.}~\bibnamefont {Brewster}}, \bibinfo {author}
  {\bibfnamefont {D.~S.}\ \bibnamefont {Wisbey}}, \ and\ \bibinfo {author}
  {\bibfnamefont {K.~W.}\ \bibnamefont {Murch}},\ }\bibfield  {title} {\enquote
  {\bibinfo {title} {Optical direct write of dolan--niemeyer-bridge junctions
  for transmon qubits},}\ }\href {\doibase 10.1063/5.0060246} {\bibfield
  {journal} {\bibinfo  {journal} {Applied Physics Letters}\ }\textbf {\bibinfo
  {volume} {119}},\ \bibinfo {pages} {062601} (\bibinfo {year} {2021})},\
  \Eprint {http://arxiv.org/abs/https://doi.org/10.1063/5.0060246}
  {https://doi.org/10.1063/5.0060246} \BibitemShut {NoStop}%
\bibitem [{\citenamefont {Halbritter}(1987)}]{Halbritter1987}%
  \BibitemOpen
  \bibfield  {author} {\bibinfo {author} {\bibfnamefont {J.}~\bibnamefont
  {Halbritter}},\ }\bibfield  {title} {\enquote {\bibinfo {title} {{On the
  oxidation and on the superconductivity of niobium}},}\ }\href {\doibase
  10.1007/BF00615201} {\bibfield  {journal} {\bibinfo  {journal} {Applied
  Physics A Solids and Surfaces}\ }\textbf {\bibinfo {volume} {43}},\ \bibinfo
  {pages} {1--28} (\bibinfo {year} {1987})}\BibitemShut {NoStop}%
\bibitem [{\citenamefont {Nersisyan}\ \emph {et~al.}(2019)\citenamefont
  {Nersisyan}, \citenamefont {Sete}, \citenamefont {Stanwyck}, \citenamefont
  {Bestwick}, \citenamefont {Reagor}, \citenamefont {Poletto}, \citenamefont
  {Alidoust}, \citenamefont {Manenti}, \citenamefont {Renzas}, \citenamefont
  {Bui}, \citenamefont {Vu}, \citenamefont {Whyland},\ and\ \citenamefont
  {Mohan}}]{Nersisyan2019}%
  \BibitemOpen
  \bibfield  {author} {\bibinfo {author} {\bibfnamefont {A.}~\bibnamefont
  {Nersisyan}}, \bibinfo {author} {\bibfnamefont {E.~A.}\ \bibnamefont {Sete}},
  \bibinfo {author} {\bibfnamefont {S.}~\bibnamefont {Stanwyck}}, \bibinfo
  {author} {\bibfnamefont {A.}~\bibnamefont {Bestwick}}, \bibinfo {author}
  {\bibfnamefont {M.}~\bibnamefont {Reagor}}, \bibinfo {author} {\bibfnamefont
  {S.}~\bibnamefont {Poletto}}, \bibinfo {author} {\bibfnamefont
  {N.}~\bibnamefont {Alidoust}}, \bibinfo {author} {\bibfnamefont
  {R.}~\bibnamefont {Manenti}}, \bibinfo {author} {\bibfnamefont
  {R.}~\bibnamefont {Renzas}}, \bibinfo {author} {\bibfnamefont {C.-V.}\
  \bibnamefont {Bui}}, \bibinfo {author} {\bibfnamefont {K.}~\bibnamefont
  {Vu}}, \bibinfo {author} {\bibfnamefont {T.}~\bibnamefont {Whyland}}, \ and\
  \bibinfo {author} {\bibfnamefont {Y.}~\bibnamefont {Mohan}},\ }\bibfield
  {title} {\enquote {\bibinfo {title} {Manufacturing low dissipation
  superconducting quantum processors},}\ }in\ \href {\doibase
  10.1109/iedm19573.2019.8993458} {\emph {\bibinfo {booktitle} {2019 {IEEE}
  International Electron Devices Meeting ({IEDM})}}}\ (\bibinfo  {publisher}
  {{IEEE}},\ \bibinfo {year} {2019})\BibitemShut {NoStop}%
\bibitem [{\citenamefont {Blok}\ \emph {et~al.}(2021)\citenamefont {Blok},
  \citenamefont {Ramasesh}, \citenamefont {Schuster}, \citenamefont {O'Brien},
  \citenamefont {Kreikebaum}, \citenamefont {Dahlen}, \citenamefont {Morvan},
  \citenamefont {Yoshida}, \citenamefont {Yao},\ and\ \citenamefont
  {Siddiqi}}]{Blok2021}%
  \BibitemOpen
  \bibfield  {author} {\bibinfo {author} {\bibfnamefont {M.~S.}\ \bibnamefont
  {Blok}}, \bibinfo {author} {\bibfnamefont {V.~V.}\ \bibnamefont {Ramasesh}},
  \bibinfo {author} {\bibfnamefont {T.}~\bibnamefont {Schuster}}, \bibinfo
  {author} {\bibfnamefont {K.}~\bibnamefont {O'Brien}}, \bibinfo {author}
  {\bibfnamefont {J.~M.}\ \bibnamefont {Kreikebaum}}, \bibinfo {author}
  {\bibfnamefont {D.}~\bibnamefont {Dahlen}}, \bibinfo {author} {\bibfnamefont
  {A.}~\bibnamefont {Morvan}}, \bibinfo {author} {\bibfnamefont
  {B.}~\bibnamefont {Yoshida}}, \bibinfo {author} {\bibfnamefont {N.~Y.}\
  \bibnamefont {Yao}}, \ and\ \bibinfo {author} {\bibfnamefont
  {I.}~\bibnamefont {Siddiqi}},\ }\bibfield  {title} {\enquote {\bibinfo
  {title} {{Quantum Information Scrambling on a Superconducting Qutrit
  Processor}},}\ }\href {\doibase 10.1103/PhysRevX.11.021010} {\bibfield
  {journal} {\bibinfo  {journal} {Physical Review X}\ }\textbf {\bibinfo
  {volume} {11}},\ \bibinfo {pages} {21010} (\bibinfo {year} {2021})},\ \Eprint
  {http://arxiv.org/abs/2003.03307} {arXiv:2003.03307} \BibitemShut {NoStop}%
\bibitem [{\citenamefont {Bach}\ \emph {et~al.}(2006)\citenamefont {Bach},
  \citenamefont {St{\"{o}}rmer}, \citenamefont {Schneider}, \citenamefont
  {Gerthsen},\ and\ \citenamefont {Verbeeck}}]{Bach2006}%
  \BibitemOpen
  \bibfield  {author} {\bibinfo {author} {\bibfnamefont {D.}~\bibnamefont
  {Bach}}, \bibinfo {author} {\bibfnamefont {H.}~\bibnamefont {St{\"{o}}rmer}},
  \bibinfo {author} {\bibfnamefont {R.}~\bibnamefont {Schneider}}, \bibinfo
  {author} {\bibfnamefont {D.}~\bibnamefont {Gerthsen}}, \ and\ \bibinfo
  {author} {\bibfnamefont {J.}~\bibnamefont {Verbeeck}},\ }\bibfield  {title}
  {\enquote {\bibinfo {title} {{EELS investigations of different niobium oxide
  phases}},}\ }\href {\doibase 10.1017/S1431927606060521} {\bibfield  {journal}
  {\bibinfo  {journal} {Microscopy and Microanalysis}\ } (\bibinfo {year}
  {2006}),\ 10.1017/S1431927606060521}\BibitemShut {NoStop}%
\bibitem [{\citenamefont {Bach}(2009)}]{Bach2009_Thesis}%
  \BibitemOpen
  \bibfield  {author} {\bibinfo {author} {\bibfnamefont {D.}~\bibnamefont
  {Bach}},\ }\emph {\bibinfo {title} {EELS investigations of stoichiometric
  niobium oxides and niobium-based capacitors}},\ \href {\doibase
  10.5445/IR/1000012945} {Ph.D. thesis},\ \bibinfo  {school} {{Universität
  Karlsruhe (TH)}} (\bibinfo {year} {2009})\BibitemShut {NoStop}%
\bibitem [{\citenamefont {Alto{\'{e}}}\ \emph {et~al.}(2020)\citenamefont
  {Alto{\'{e}}}, \citenamefont {Banerjee}, \citenamefont {Berk}, \citenamefont
  {Hajr}, \citenamefont {Schwartzberg}, \citenamefont {Song}, \citenamefont
  {Ghadeer}, \citenamefont {Aloni}, \citenamefont {Elowson}, \citenamefont
  {Kreikebaum}, \citenamefont {Wong}, \citenamefont {Griffin}, \citenamefont
  {Rao}, \citenamefont {Weber-Bargioni}, \citenamefont {Minor}, \citenamefont
  {Santiago}, \citenamefont {Cabrini}, \citenamefont {Siddiqi},\ and\
  \citenamefont {Ogletree}}]{Altoe2020}%
  \BibitemOpen
  \bibfield  {author} {\bibinfo {author} {\bibfnamefont {M.~V.~P.}\
  \bibnamefont {Alto{\'{e}}}}, \bibinfo {author} {\bibfnamefont
  {A.}~\bibnamefont {Banerjee}}, \bibinfo {author} {\bibfnamefont
  {C.}~\bibnamefont {Berk}}, \bibinfo {author} {\bibfnamefont {A.}~\bibnamefont
  {Hajr}}, \bibinfo {author} {\bibfnamefont {A.}~\bibnamefont {Schwartzberg}},
  \bibinfo {author} {\bibfnamefont {C.}~\bibnamefont {Song}}, \bibinfo {author}
  {\bibfnamefont {M.~A.}\ \bibnamefont {Ghadeer}}, \bibinfo {author}
  {\bibfnamefont {S.}~\bibnamefont {Aloni}}, \bibinfo {author} {\bibfnamefont
  {M.~J.}\ \bibnamefont {Elowson}}, \bibinfo {author} {\bibfnamefont {J.~M.}\
  \bibnamefont {Kreikebaum}}, \bibinfo {author} {\bibfnamefont {E.~K.}\
  \bibnamefont {Wong}}, \bibinfo {author} {\bibfnamefont {S.}~\bibnamefont
  {Griffin}}, \bibinfo {author} {\bibfnamefont {S.}~\bibnamefont {Rao}},
  \bibinfo {author} {\bibfnamefont {A.}~\bibnamefont {Weber-Bargioni}},
  \bibinfo {author} {\bibfnamefont {A.~M.}\ \bibnamefont {Minor}}, \bibinfo
  {author} {\bibfnamefont {D.~I.}\ \bibnamefont {Santiago}}, \bibinfo {author}
  {\bibfnamefont {S.}~\bibnamefont {Cabrini}}, \bibinfo {author} {\bibfnamefont
  {I.}~\bibnamefont {Siddiqi}}, \ and\ \bibinfo {author} {\bibfnamefont
  {D.~F.}\ \bibnamefont {Ogletree}},\ }\bibfield  {title} {\enquote {\bibinfo
  {title} {{Localization and reduction of superconducting quantum coherent
  circuit losses}},}\ }\href {http://arxiv.org/abs/2012.07604} {\bibfield
  {journal} {\bibinfo  {journal} {arXiv preprints}\ ,\ \bibinfo {pages}
  {1--20}} (\bibinfo {year} {2020})},\ \Eprint
  {http://arxiv.org/abs/2012.07604} {arXiv:2012.07604} \BibitemShut {NoStop}%
\bibitem [{\citenamefont {Verjauw}\ \emph {et~al.}(2021)\citenamefont
  {Verjauw}, \citenamefont {Poto\ifmmode~\check{c}\else \v{c}\fi{}nik},
  \citenamefont {Mongillo}, \citenamefont {Acharya}, \citenamefont
  {Mohiyaddin}, \citenamefont {Simion}, \citenamefont {Pacco}, \citenamefont
  {Ivanov}, \citenamefont {Wan}, \citenamefont {Vanleenhove}, \citenamefont
  {Souriau}, \citenamefont {Jussot}, \citenamefont {Thiam}, \citenamefont
  {Swerts}, \citenamefont {Piao}, \citenamefont {Couet}, \citenamefont {Heyns},
  \citenamefont {Govoreanu},\ and\ \citenamefont {Radu}}]{Verjauw2021}%
  \BibitemOpen
  \bibfield  {author} {\bibinfo {author} {\bibfnamefont {J.}~\bibnamefont
  {Verjauw}}, \bibinfo {author} {\bibfnamefont {A.}~\bibnamefont
  {Poto\ifmmode~\check{c}\else \v{c}\fi{}nik}}, \bibinfo {author}
  {\bibfnamefont {M.}~\bibnamefont {Mongillo}}, \bibinfo {author}
  {\bibfnamefont {R.}~\bibnamefont {Acharya}}, \bibinfo {author} {\bibfnamefont
  {F.}~\bibnamefont {Mohiyaddin}}, \bibinfo {author} {\bibfnamefont
  {G.}~\bibnamefont {Simion}}, \bibinfo {author} {\bibfnamefont
  {A.}~\bibnamefont {Pacco}}, \bibinfo {author} {\bibfnamefont
  {T.}~\bibnamefont {Ivanov}}, \bibinfo {author} {\bibfnamefont
  {D.}~\bibnamefont {Wan}}, \bibinfo {author} {\bibfnamefont {A.}~\bibnamefont
  {Vanleenhove}}, \bibinfo {author} {\bibfnamefont {L.}~\bibnamefont
  {Souriau}}, \bibinfo {author} {\bibfnamefont {J.}~\bibnamefont {Jussot}},
  \bibinfo {author} {\bibfnamefont {A.}~\bibnamefont {Thiam}}, \bibinfo
  {author} {\bibfnamefont {J.}~\bibnamefont {Swerts}}, \bibinfo {author}
  {\bibfnamefont {X.}~\bibnamefont {Piao}}, \bibinfo {author} {\bibfnamefont
  {S.}~\bibnamefont {Couet}}, \bibinfo {author} {\bibfnamefont
  {M.}~\bibnamefont {Heyns}}, \bibinfo {author} {\bibfnamefont
  {B.}~\bibnamefont {Govoreanu}}, \ and\ \bibinfo {author} {\bibfnamefont
  {I.}~\bibnamefont {Radu}},\ }\bibfield  {title} {\enquote {\bibinfo {title}
  {Investigation of microwave loss induced by oxide regrowth in high-q niobium
  resonators},}\ }\href {\doibase 10.1103/PhysRevApplied.16.014018} {\bibfield
  {journal} {\bibinfo  {journal} {Phys. Rev. Applied}\ }\textbf {\bibinfo
  {volume} {16}},\ \bibinfo {pages} {014018} (\bibinfo {year}
  {2021})}\BibitemShut {NoStop}%
\bibitem [{\citenamefont {Siemers}\ \emph {et~al.}(2014)\citenamefont
  {Siemers}, \citenamefont {Pflug}, \citenamefont {Melzig}, \citenamefont
  {Gehrke}, \citenamefont {Weimar},\ and\ \citenamefont
  {Szyszka}}]{Siemers2014}%
  \BibitemOpen
  \bibfield  {author} {\bibinfo {author} {\bibfnamefont {M.}~\bibnamefont
  {Siemers}}, \bibinfo {author} {\bibfnamefont {A.}~\bibnamefont {Pflug}},
  \bibinfo {author} {\bibfnamefont {T.}~\bibnamefont {Melzig}}, \bibinfo
  {author} {\bibfnamefont {K.}~\bibnamefont {Gehrke}}, \bibinfo {author}
  {\bibfnamefont {A.}~\bibnamefont {Weimar}}, \ and\ \bibinfo {author}
  {\bibfnamefont {B.}~\bibnamefont {Szyszka}},\ }\bibfield  {title} {\enquote
  {\bibinfo {title} {{Model based investigation of Ar+ ion damage in DC
  magnetron sputtering}},}\ }\href {\doibase 10.1016/j.surfcoat.2013.09.025}
  {\bibfield  {journal} {\bibinfo  {journal} {Surface and Coatings Technology}\
  } (\bibinfo {year} {2014}),\ 10.1016/j.surfcoat.2013.09.025}\BibitemShut
  {NoStop}%
\bibitem [{\citenamefont {D'Heurle}(1970)}]{DHeurle1970}%
  \BibitemOpen
  \bibfield  {author} {\bibinfo {author} {\bibfnamefont {F.~M.}\ \bibnamefont
  {D'Heurle}},\ }\bibfield  {title} {\enquote {\bibinfo {title} {{Aluminum
  films deposited by rf sputtering}},}\ }\href {\doibase 10.1007/BF02811600}
  {\bibfield  {journal} {\bibinfo  {journal} {Metallurgical and Materials
  Transactions B}\ } (\bibinfo {year} {1970}),\ 10.1007/BF02811600}\BibitemShut
  {NoStop}%
\bibitem [{\citenamefont {Morohashi}\ \emph {et~al.}(2001)\citenamefont
  {Morohashi}, \citenamefont {Takeda}, \citenamefont {Tsujimura}, \citenamefont
  {Kawanishi}, \citenamefont {Harada}, \citenamefont {Maekawa}, \citenamefont
  {Nakayama},\ and\ \citenamefont {Noguchi}}]{Morohashi2001}%
  \BibitemOpen
  \bibfield  {author} {\bibinfo {author} {\bibfnamefont {S.}~\bibnamefont
  {Morohashi}}, \bibinfo {author} {\bibfnamefont {N.}~\bibnamefont {Takeda}},
  \bibinfo {author} {\bibfnamefont {S.}~\bibnamefont {Tsujimura}}, \bibinfo
  {author} {\bibfnamefont {M.}~\bibnamefont {Kawanishi}}, \bibinfo {author}
  {\bibfnamefont {K.}~\bibnamefont {Harada}}, \bibinfo {author} {\bibfnamefont
  {S.}~\bibnamefont {Maekawa}}, \bibinfo {author} {\bibfnamefont
  {N.}~\bibnamefont {Nakayama}}, \ and\ \bibinfo {author} {\bibfnamefont
  {T.}~\bibnamefont {Noguchi}},\ }\bibfield  {title} {\enquote {\bibinfo
  {title} {{Characteristics of superconducting Nb layer fabricated using
  high-vacuum electron beam evaporation}},}\ }\href {\doibase
  10.1143/jjap.40.576} {\bibfield  {journal} {\bibinfo  {journal} {Japanese
  Journal of Applied Physics, Part 1: Regular Papers and Short Notes and Review
  Papers}\ }\textbf {\bibinfo {volume} {40}},\ \bibinfo {pages} {576--579}
  (\bibinfo {year} {2001})}\BibitemShut {NoStop}%
\bibitem [{\citenamefont {Pappas}\ \emph {et~al.}(2011)\citenamefont {Pappas},
  \citenamefont {Vissers}, \citenamefont {Wisbey}, \citenamefont {Kline},\ and\
  \citenamefont {Gao}}]{Pappas2011}%
  \BibitemOpen
  \bibfield  {author} {\bibinfo {author} {\bibfnamefont {D.~P.}\ \bibnamefont
  {Pappas}}, \bibinfo {author} {\bibfnamefont {M.~R.}\ \bibnamefont {Vissers}},
  \bibinfo {author} {\bibfnamefont {D.~S.}\ \bibnamefont {Wisbey}}, \bibinfo
  {author} {\bibfnamefont {J.~S.}\ \bibnamefont {Kline}}, \ and\ \bibinfo
  {author} {\bibfnamefont {J.}~\bibnamefont {Gao}},\ }\bibfield  {title}
  {\enquote {\bibinfo {title} {{Two level system loss in superconducting
  microwave resonators}},}\ }in\ \href {\doibase 10.1109/TASC.2010.2097578}
  {\emph {\bibinfo {booktitle} {IEEE Transactions on Applied
  Superconductivity}}}\ (\bibinfo {year} {2011})\BibitemShut {NoStop}%
\bibitem [{\citenamefont {McRae}\ \emph {et~al.}(2020)\citenamefont {McRae},
  \citenamefont {Wang}, \citenamefont {Gao}, \citenamefont {Vissers},
  \citenamefont {Brecht}, \citenamefont {Dunsworth}, \citenamefont {Pappas},\
  and\ \citenamefont {Mutus}}]{McRae2020}%
  \BibitemOpen
  \bibfield  {author} {\bibinfo {author} {\bibfnamefont {C.~R.~H.}\
  \bibnamefont {McRae}}, \bibinfo {author} {\bibfnamefont {H.}~\bibnamefont
  {Wang}}, \bibinfo {author} {\bibfnamefont {J.}~\bibnamefont {Gao}}, \bibinfo
  {author} {\bibfnamefont {M.~R.}\ \bibnamefont {Vissers}}, \bibinfo {author}
  {\bibfnamefont {T.}~\bibnamefont {Brecht}}, \bibinfo {author} {\bibfnamefont
  {A.}~\bibnamefont {Dunsworth}}, \bibinfo {author} {\bibfnamefont {D.~P.}\
  \bibnamefont {Pappas}}, \ and\ \bibinfo {author} {\bibfnamefont
  {J.}~\bibnamefont {Mutus}},\ }\bibfield  {title} {\enquote {\bibinfo {title}
  {Materials loss measurements using superconducting microwave resonators},}\
  }\href {\doibase 10.1063/5.0017378} {\bibfield  {journal} {\bibinfo
  {journal} {Review of Scientific Instruments}\ }\textbf {\bibinfo {volume}
  {91}},\ \bibinfo {pages} {091101} (\bibinfo {year} {2020})},\ \Eprint
  {http://arxiv.org/abs/https://doi.org/10.1063/5.0017378}
  {https://doi.org/10.1063/5.0017378} \BibitemShut {NoStop}%
\bibitem [{\citenamefont {Chang}\ \emph {et~al.}(2013)\citenamefont {Chang},
  \citenamefont {Vissers}, \citenamefont {Córcoles}, \citenamefont {Sandberg},
  \citenamefont {Gao}, \citenamefont {Abraham}, \citenamefont {Chow},
  \citenamefont {Gambetta}, \citenamefont {Beth~Rothwell}, \citenamefont
  {Keefe}, \citenamefont {Steffen},\ and\ \citenamefont {Pappas}}]{Chang_2013}%
  \BibitemOpen
  \bibfield  {author} {\bibinfo {author} {\bibfnamefont {J.~B.}\ \bibnamefont
  {Chang}}, \bibinfo {author} {\bibfnamefont {M.~R.}\ \bibnamefont {Vissers}},
  \bibinfo {author} {\bibfnamefont {A.~D.}\ \bibnamefont {Córcoles}}, \bibinfo
  {author} {\bibfnamefont {M.}~\bibnamefont {Sandberg}}, \bibinfo {author}
  {\bibfnamefont {J.}~\bibnamefont {Gao}}, \bibinfo {author} {\bibfnamefont
  {D.~W.}\ \bibnamefont {Abraham}}, \bibinfo {author} {\bibfnamefont {J.~M.}\
  \bibnamefont {Chow}}, \bibinfo {author} {\bibfnamefont {J.~M.}\ \bibnamefont
  {Gambetta}}, \bibinfo {author} {\bibfnamefont {M.}~\bibnamefont
  {Beth~Rothwell}}, \bibinfo {author} {\bibfnamefont {G.~A.}\ \bibnamefont
  {Keefe}}, \bibinfo {author} {\bibfnamefont {M.}~\bibnamefont {Steffen}}, \
  and\ \bibinfo {author} {\bibfnamefont {D.~P.}\ \bibnamefont {Pappas}},\
  }\bibfield  {title} {\enquote {\bibinfo {title} {Improved superconducting
  qubit coherence using titanium nitride},}\ }\href {\doibase
  10.1063/1.4813269} {\bibfield  {journal} {\bibinfo  {journal} {Applied
  Physics Letters}\ }\textbf {\bibinfo {volume} {103}},\ \bibinfo {pages}
  {012602} (\bibinfo {year} {2013})},\ \Eprint
  {http://arxiv.org/abs/https://doi.org/10.1063/1.4813269}
  {https://doi.org/10.1063/1.4813269} \BibitemShut {NoStop}%
\bibitem [{\citenamefont {Zeng}\ \emph {et~al.}(2015)\citenamefont {Zeng},
  \citenamefont {Krantz}, \citenamefont {Nik}, \citenamefont {Delsing},\ and\
  \citenamefont {Olsson}}]{Zeng2015}%
  \BibitemOpen
  \bibfield  {author} {\bibinfo {author} {\bibfnamefont {L.~J.}\ \bibnamefont
  {Zeng}}, \bibinfo {author} {\bibfnamefont {P.}~\bibnamefont {Krantz}},
  \bibinfo {author} {\bibfnamefont {S.}~\bibnamefont {Nik}}, \bibinfo {author}
  {\bibfnamefont {P.}~\bibnamefont {Delsing}}, \ and\ \bibinfo {author}
  {\bibfnamefont {E.}~\bibnamefont {Olsson}},\ }\bibfield  {title} {\enquote
  {\bibinfo {title} {{The atomic details of the interfacial interaction between
  the bottom electrode of Al/AlOx/Al Josephson junctions and HF-treated Si
  substrates}},}\ }\href {\doibase 10.1063/1.4919224} {\bibfield  {journal}
  {\bibinfo  {journal} {Journal of Applied Physics}\ } (\bibinfo {year}
  {2015}),\ 10.1063/1.4919224}\BibitemShut {NoStop}%
\bibitem [{\citenamefont {Premkumar}\ \emph {et~al.}(2021)\citenamefont
  {Premkumar}, \citenamefont {Weiland}, \citenamefont {Hwang}, \citenamefont
  {J{\"{a}}ck}, \citenamefont {Place}, \citenamefont {Waluyo}, \citenamefont
  {Hunt}, \citenamefont {Bisogni}, \citenamefont {Pelliciari}, \citenamefont
  {Barbour}, \citenamefont {Miller}, \citenamefont {Russo}, \citenamefont
  {Camino}, \citenamefont {Kisslinger}, \citenamefont {Tong}, \citenamefont
  {Hybertsen}, \citenamefont {Houck},\ and\ \citenamefont
  {Jarrige}}]{Premkumar2021}%
  \BibitemOpen
  \bibfield  {author} {\bibinfo {author} {\bibfnamefont {A.}~\bibnamefont
  {Premkumar}}, \bibinfo {author} {\bibfnamefont {C.}~\bibnamefont {Weiland}},
  \bibinfo {author} {\bibfnamefont {S.}~\bibnamefont {Hwang}}, \bibinfo
  {author} {\bibfnamefont {B.}~\bibnamefont {J{\"{a}}ck}}, \bibinfo {author}
  {\bibfnamefont {A.~P.~M.}\ \bibnamefont {Place}}, \bibinfo {author}
  {\bibfnamefont {I.}~\bibnamefont {Waluyo}}, \bibinfo {author} {\bibfnamefont
  {A.}~\bibnamefont {Hunt}}, \bibinfo {author} {\bibfnamefont {V.}~\bibnamefont
  {Bisogni}}, \bibinfo {author} {\bibfnamefont {J.}~\bibnamefont {Pelliciari}},
  \bibinfo {author} {\bibfnamefont {A.}~\bibnamefont {Barbour}}, \bibinfo
  {author} {\bibfnamefont {M.~S.}\ \bibnamefont {Miller}}, \bibinfo {author}
  {\bibfnamefont {P.}~\bibnamefont {Russo}}, \bibinfo {author} {\bibfnamefont
  {F.}~\bibnamefont {Camino}}, \bibinfo {author} {\bibfnamefont
  {K.}~\bibnamefont {Kisslinger}}, \bibinfo {author} {\bibfnamefont
  {X.}~\bibnamefont {Tong}}, \bibinfo {author} {\bibfnamefont {M.~S.}\
  \bibnamefont {Hybertsen}}, \bibinfo {author} {\bibfnamefont {A.~A.}\
  \bibnamefont {Houck}}, \ and\ \bibinfo {author} {\bibfnamefont
  {I.}~\bibnamefont {Jarrige}},\ }\bibfield  {title} {\enquote {\bibinfo
  {title} {{Microscopic relaxation channels in materials for superconducting
  qubits}},}\ }\href {\doibase 10.1038/s43246-021-00174-7} {\bibfield
  {journal} {\bibinfo  {journal} {Communications Materials}\ }\textbf {\bibinfo
  {volume} {2}} (\bibinfo {year} {2021}),\
  10.1038/s43246-021-00174-7}\BibitemShut {NoStop}%
\bibitem [{\citenamefont {Bonin}\ and\ \citenamefont {Safa}(1991)}]{Bonin1991}%
  \BibitemOpen
  \bibfield  {author} {\bibinfo {author} {\bibfnamefont {B.}~\bibnamefont
  {Bonin}}\ and\ \bibinfo {author} {\bibfnamefont {H.}~\bibnamefont {Safa}},\
  }\bibfield  {title} {\enquote {\bibinfo {title} {{Power dissipation at high
  fields in granular RF superconductivity}},}\ }\href {\doibase
  10.1088/0953-2048/4/6/008} {\bibfield  {journal} {\bibinfo  {journal}
  {Superconductor Science and Technology}\ }\textbf {\bibinfo {volume} {4}},\
  \bibinfo {pages} {257--261} (\bibinfo {year} {1991})}\BibitemShut {NoStop}%
\bibitem [{\citenamefont {Newville}\ \emph {et~al.}(2014)\citenamefont
  {Newville}, \citenamefont {Ingargiola}, \citenamefont {Stensitzki},\ and\
  \citenamefont {Allen}}]{Newville2014}%
  \BibitemOpen
  \bibfield  {author} {\bibinfo {author} {\bibfnamefont {M.}~\bibnamefont
  {Newville}}, \bibinfo {author} {\bibfnamefont {A.}~\bibnamefont
  {Ingargiola}}, \bibinfo {author} {\bibfnamefont {T.}~\bibnamefont
  {Stensitzki}}, \ and\ \bibinfo {author} {\bibfnamefont {D.~B.}\ \bibnamefont
  {Allen}},\ }\bibfield  {title} {\enquote {\bibinfo {title} {{LMFIT:
  Non-Linear Least-Square Minimization and Curve-Fitting for Python}},}\
  }\href@noop {} {\bibfield  {journal} {\bibinfo  {journal} {Zenodo}\ }
  (\bibinfo {year} {2014})}\BibitemShut {NoStop}%
\bibitem [{\citenamefont {Hulm}\ \emph {et~al.}(1972)\citenamefont {Hulm},
  \citenamefont {Jones}, \citenamefont {Hein},\ and\ \citenamefont
  {Gibson}}]{Hulm1972}%
  \BibitemOpen
  \bibfield  {author} {\bibinfo {author} {\bibfnamefont {J.~K.}\ \bibnamefont
  {Hulm}}, \bibinfo {author} {\bibfnamefont {C.~K.}\ \bibnamefont {Jones}},
  \bibinfo {author} {\bibfnamefont {R.~A.}\ \bibnamefont {Hein}}, \ and\
  \bibinfo {author} {\bibfnamefont {J.~W.}\ \bibnamefont {Gibson}},\ }\bibfield
   {title} {\enquote {\bibinfo {title} {{Superconductivity in the TiO and NbO
  systems}},}\ }\href {\doibase 10.1007/BF00660068} {\bibfield  {journal}
  {\bibinfo  {journal} {Journal of Low Temperature Physics}\ } (\bibinfo {year}
  {1972}),\ 10.1007/BF00660068}\BibitemShut {NoStop}%
\bibitem [{\citenamefont {Soares}\ \emph {et~al.}(2011)\citenamefont {Soares},
  \citenamefont {Leite}, \citenamefont {Nico}, \citenamefont {Peres},
  \citenamefont {Fernandes}, \citenamefont {Gra{\c{c}}a}, \citenamefont
  {Matos}, \citenamefont {Monteiro}, \citenamefont {Monteiro},\ and\
  \citenamefont {Costa}}]{Soares2011}%
  \BibitemOpen
  \bibfield  {author} {\bibinfo {author} {\bibfnamefont {M.~R.}\ \bibnamefont
  {Soares}}, \bibinfo {author} {\bibfnamefont {S.}~\bibnamefont {Leite}},
  \bibinfo {author} {\bibfnamefont {C.}~\bibnamefont {Nico}}, \bibinfo {author}
  {\bibfnamefont {M.}~\bibnamefont {Peres}}, \bibinfo {author} {\bibfnamefont
  {A.~J.}\ \bibnamefont {Fernandes}}, \bibinfo {author} {\bibfnamefont {M.~P.}\
  \bibnamefont {Gra{\c{c}}a}}, \bibinfo {author} {\bibfnamefont
  {M.}~\bibnamefont {Matos}}, \bibinfo {author} {\bibfnamefont
  {R.}~\bibnamefont {Monteiro}}, \bibinfo {author} {\bibfnamefont
  {T.}~\bibnamefont {Monteiro}}, \ and\ \bibinfo {author} {\bibfnamefont
  {F.~M.}\ \bibnamefont {Costa}},\ }\bibfield  {title} {\enquote {\bibinfo
  {title} {{Effect of processing method on physical properties of Nb2O5}},}\
  }\href {\doibase 10.1016/j.jeurceramsoc.2010.10.024} {\bibfield  {journal}
  {\bibinfo  {journal} {Journal of the European Ceramic Society}\ } (\bibinfo
  {year} {2011}),\ 10.1016/j.jeurceramsoc.2010.10.024}\BibitemShut {NoStop}%
\bibitem [{\citenamefont {Nico}, \citenamefont {Monteiro},\ and\ \citenamefont
  {Gra{\c{c}}a}(2016)}]{Nico2016}%
  \BibitemOpen
  \bibfield  {author} {\bibinfo {author} {\bibfnamefont {C.}~\bibnamefont
  {Nico}}, \bibinfo {author} {\bibfnamefont {T.}~\bibnamefont {Monteiro}}, \
  and\ \bibinfo {author} {\bibfnamefont {M.~P.}\ \bibnamefont {Gra{\c{c}}a}},\
  }\href {\doibase 10.1016/j.pmatsci.2016.02.001} {\enquote {\bibinfo {title}
  {{Niobium oxides and niobates physical properties: Review and prospects}},}\
  } (\bibinfo {year} {2016})\BibitemShut {NoStop}%
\bibitem [{\citenamefont {Zmuidzinas}(2012)}]{Zmuidzinas2012}%
  \BibitemOpen
  \bibfield  {author} {\bibinfo {author} {\bibfnamefont {J.}~\bibnamefont
  {Zmuidzinas}},\ }\bibfield  {title} {\enquote {\bibinfo {title}
  {Superconducting microresonators: Physics and applications},}\ }\href
  {\doibase 10.1146/annurev-conmatphys-020911-125022} {\bibfield  {journal}
  {\bibinfo  {journal} {Annual Review of Condensed Matter Physics}\ }\textbf
  {\bibinfo {volume} {3}},\ \bibinfo {pages} {169--214} (\bibinfo {year}
  {2012})},\ \Eprint
  {http://arxiv.org/abs/https://doi.org/10.1146/annurev-conmatphys-020911-125022}
  {https://doi.org/10.1146/annurev-conmatphys-020911-125022} \BibitemShut
  {NoStop}%
\bibitem [{\citenamefont {Wisbey}\ \emph {et~al.}(2019)\citenamefont {Wisbey},
  \citenamefont {Vissers}, \citenamefont {Gao}, \citenamefont {Kline},
  \citenamefont {Sandberg}, \citenamefont {Weides}, \citenamefont {Paquette},
  \citenamefont {Karki}, \citenamefont {Brewster}, \citenamefont {Alameri},
  \citenamefont {Kuljanishvili}, \citenamefont {Caruso},\ and\ \citenamefont
  {Pappas}}]{Wisbey2019}%
  \BibitemOpen
  \bibfield  {author} {\bibinfo {author} {\bibfnamefont {D.~S.}\ \bibnamefont
  {Wisbey}}, \bibinfo {author} {\bibfnamefont {M.~R.}\ \bibnamefont {Vissers}},
  \bibinfo {author} {\bibfnamefont {J.}~\bibnamefont {Gao}}, \bibinfo {author}
  {\bibfnamefont {J.~S.}\ \bibnamefont {Kline}}, \bibinfo {author}
  {\bibfnamefont {M.~O.}\ \bibnamefont {Sandberg}}, \bibinfo {author}
  {\bibfnamefont {M.~P.}\ \bibnamefont {Weides}}, \bibinfo {author}
  {\bibfnamefont {M.~M.}\ \bibnamefont {Paquette}}, \bibinfo {author}
  {\bibfnamefont {S.}~\bibnamefont {Karki}}, \bibinfo {author} {\bibfnamefont
  {J.}~\bibnamefont {Brewster}}, \bibinfo {author} {\bibfnamefont
  {D.}~\bibnamefont {Alameri}}, \bibinfo {author} {\bibfnamefont
  {I.}~\bibnamefont {Kuljanishvili}}, \bibinfo {author} {\bibfnamefont {A.~N.}\
  \bibnamefont {Caruso}}, \ and\ \bibinfo {author} {\bibfnamefont {D.~P.}\
  \bibnamefont {Pappas}},\ }\bibfield  {title} {\enquote {\bibinfo {title}
  {Dielectric loss of boron-based dielectrics on niobium resonators},}\ }\href
  {\doibase 10.1007/s10909-019-02183-w} {\bibfield  {journal} {\bibinfo
  {journal} {Journal of Low Temperature Physics}\ }\textbf {\bibinfo {volume}
  {195}},\ \bibinfo {pages} {474--486} (\bibinfo {year} {2019})}\BibitemShut
  {NoStop}%
\bibitem [{\citenamefont {Gao}(2008)}]{Gao2008_thesis}%
  \BibitemOpen
  \bibfield  {author} {\bibinfo {author} {\bibfnamefont {J.}~\bibnamefont
  {Gao}},\ }\emph {\bibinfo {title} {The Physics of Superconducting Microwave
  Resonators}},\ \href@noop {} {Ph.D. thesis},\ \bibinfo  {school} {California
  Institute of Technology} (\bibinfo {year} {2008})\BibitemShut {NoStop}%
\bibitem [{\citenamefont {Burnett}, \citenamefont {Faoro},\ and\ \citenamefont
  {Lindström}(2016)}]{Burnett_2016}%
  \BibitemOpen
  \bibfield  {author} {\bibinfo {author} {\bibfnamefont {J.}~\bibnamefont
  {Burnett}}, \bibinfo {author} {\bibfnamefont {L.}~\bibnamefont {Faoro}}, \
  and\ \bibinfo {author} {\bibfnamefont {T.}~\bibnamefont {Lindström}},\
  }\bibfield  {title} {\enquote {\bibinfo {title} {Analysis of high quality
  superconducting resonators: consequences for {TLS} properties in amorphous
  oxides},}\ }\href {\doibase 10.1088/0953-2048/29/4/044008} {\bibfield
  {journal} {\bibinfo  {journal} {Superconductor Science and Technology}\
  }\textbf {\bibinfo {volume} {29}},\ \bibinfo {pages} {044008} (\bibinfo
  {year} {2016})}\BibitemShut {NoStop}%
\bibitem [{\citenamefont {Melville}\ \emph {et~al.}(2020)\citenamefont
  {Melville}, \citenamefont {Calusine}, \citenamefont {Woods}, \citenamefont
  {Serniak}, \citenamefont {Golden}, \citenamefont {Niedzielski}, \citenamefont
  {Kim}, \citenamefont {Sevi}, \citenamefont {Yoder}, \citenamefont {Dauler},\
  and\ \citenamefont {Oliver}}]{Melville2020}%
  \BibitemOpen
  \bibfield  {author} {\bibinfo {author} {\bibfnamefont {A.}~\bibnamefont
  {Melville}}, \bibinfo {author} {\bibfnamefont {G.}~\bibnamefont {Calusine}},
  \bibinfo {author} {\bibfnamefont {W.}~\bibnamefont {Woods}}, \bibinfo
  {author} {\bibfnamefont {K.}~\bibnamefont {Serniak}}, \bibinfo {author}
  {\bibfnamefont {E.}~\bibnamefont {Golden}}, \bibinfo {author} {\bibfnamefont
  {B.~M.}\ \bibnamefont {Niedzielski}}, \bibinfo {author} {\bibfnamefont
  {D.~K.}\ \bibnamefont {Kim}}, \bibinfo {author} {\bibfnamefont
  {A.}~\bibnamefont {Sevi}}, \bibinfo {author} {\bibfnamefont {J.~L.}\
  \bibnamefont {Yoder}}, \bibinfo {author} {\bibfnamefont {E.~A.}\ \bibnamefont
  {Dauler}}, \ and\ \bibinfo {author} {\bibfnamefont {W.~D.}\ \bibnamefont
  {Oliver}},\ }\bibfield  {title} {\enquote {\bibinfo {title} {Comparison of
  dielectric loss in titanium nitride and aluminum superconducting
  resonators},}\ }\href {\doibase 10.1063/5.0021950} {\bibfield  {journal}
  {\bibinfo  {journal} {Applied Physics Letters}\ }\textbf {\bibinfo {volume}
  {117}},\ \bibinfo {pages} {124004} (\bibinfo {year} {2020})},\ \Eprint
  {http://arxiv.org/abs/https://doi.org/10.1063/5.0021950}
  {https://doi.org/10.1063/5.0021950} \BibitemShut {NoStop}%
\end{thebibliography}

%

\end{document}